\pdfoutput=1
\documentclass[PAPER, atlasdraft=false, cernpreprint, UKenglish, texmf, orcidlogo]{atlasdoc}
\usepackage{atlaspackage}
\usepackage{atlasbiblatex}

\usepackage{atlasphysics}
\graphicspath{{logos/}{figures/}}

\usepackage{ANA-HDBS-2021-08-PAPER-defs}
\usepackage{slashbox}
\AtlasTitle{Search for a light CP-odd Higgs boson decaying into a pair of $\tau$-leptons in proton--proton collisions at $\sqrt{s}=13\,\mathrm{TeV}$ with the ATLAS detector}

\AtlasAbstract{%
This paper reports a search for a light CP-odd scalar resonance with a mass of $20\,\mathrm{GeV}$ to $90\,\mathrm{GeV}$ in $13\,\mathrm{TeV}$ proton--proton collision data with an integrated luminosity of $140\,\mathrm{fb}^{-1}$ collected with the ATLAS detector at the Large Hadron Collider.
The analysis assumes the resonance is produced via gluon--gluon fusion and decays into a $\tau^{+}\tau^{-}$ pair which subsequently decays into a fully leptonic $\mu^{+}\nu_{\mu} \bar{\nu}_{\tau} e^{-} \bar{\nu}_{e} \nu_{\tau}$ or $e^{+}\nu_{e}\bar{\nu}_{\tau} \mu^-\bar{\nu}_{\mu}\nu_{\tau}$ final state.
No significant excess of events above the predicted Standard Model background is observed.
The results are interpreted within a flavour-aligned two-Higgs-doublet model, and a model-independent cross-section interpretation is also given.
Upper limits at $95\%$ confidence level between $3.0\,\mathrm{pb}$ and $68\,\mathrm{pb}$ are set on the cross-section for producing a CP-odd Higgs boson that decays into a $\tau^+\tau^-$ pair.
}

\AtlasRefCode{HDBS-2021-08}

\PreprintIdNumber{CERN-EP-2024-235}

\arXivId{2409.20381}

\HepDataRecord{https://www.hepdata.net/record/153948}

\AtlasJournal{JHEP}
\AtlasJournalRef{JHEP 12 (2024) 126}
\AtlasDOI{10.1007/JHEP12(2024)126}

\hypersetup{pdftitle={ATLAS document},pdfauthor={The ATLAS Collaboration}}

\begin{document}

\maketitle

\tableofcontents

\section{Introduction}
\label{sec:introduction}
%

The discovery of a Higgs boson~\cite{HIGG-2012-27,CMS-HIG-12-028} at the Large Hadron Collider (LHC)~\cite{Evans:2008zzb} has provided important insight into the mechanism of electroweak symmetry breaking. Experimental studies of the Higgs boson~\cite{HIGG-2021-23,CMS-HIG-22-001} demonstrate consistency with Standard Model (SM) predictions.
However, the discovered particle could still be part of an extended scalar sector motivated by several theoretical arguments~\cite{Djouadi:2005gj,Branco:2011iw}.
One such extension of the SM is the flavour-aligned two-Higgs-doublet model (2HDM), which introduces a second SU(2) doublet~\cite{yukawa-alignment}. The couplings of the two doublets to fermions are aligned in flavour space such that one doublet obtains Yukawa couplings like those in the SM, while the second has Yukawa couplings proportional to the former ones, which prevents flavour-changing neutral currents. The proportionality constants between the two diagonalised Yukawa matrices are the fermionic coupling parameters $\zeta_f$.
Within this model, five physical Higgs bosons arise, of which two are neutral and CP-even ($h$, $H$), one is neutral and CP-odd ($A$), and the remaining two are charged ($H^\pm$).
The model thus incorporates a Higgs boson $h$ with SM properties, while allowing the presence of an additional CP-odd $A$ boson with a mass $m_A$ smaller than the mass of the $h$ boson.
A search for a light CP-odd $A$ boson hence provides further insights into the scalar sector and may shed light on its particle multiplicity.

Further interest in the search for a light $A$ boson originates from the anomalous magnetic moment of the muon, $a_\mu$, which has been a topic of discussion because of an observed discrepancy of up to \SI{5}{$\sigma$} between the experimental measurement and SM predictions~\cite{AOYAMA20201,PhysRevLett.126.141801,PhysRevLett.131.161802}.
While the theoretical situation is unclear due to contradictory calculations of the hadronic contributions~\cite{Colangelo:2022jxc, Borsanyi_2021, Boccaletti:2024, PhysRevD.109.112002}, the apparent discrepancy motivates searches for physics beyond the SM.
The flavour-aligned 2HDM predicts additional contributions to $a_\mu$ and can therefore significantly impact the level of agreement with experimental measurements~\cite{AOYAMA20201,PhysRevD.98.035001}.
In particular, fermionic two-loop contributions of the Barr--Zee type~\cite{PhysRevLett-65-21} involving $\tau$-lepton and top-quark loops coupling to Higgs bosons and to the external photon are important~\cite{Ilisie2015,PhysRevD.63.091301,PhysRevD.64.111301,PhysRevD.64.115018,Krawczyk2002}.
However, it was shown that near-maximal contributions may be required to explain the observed value of $a_\mu$~\cite{Athron:2021iuf,PhysRevD.98.035001}.
For neutral and charged Higgs bosons with masses $m_H=m_{H^\pm}$ in the range of \SI{150}{\giga\electronvolt} to \SI{250}{\giga\electronvolt}, the $a_\mu$ contributions are maximised if the CP-odd Higgs boson $A$ is light, with a mass between \SI{5}{\giga\electronvolt} and \SI{105}{\giga\electronvolt}.
While loop contributions to $a_\mu$ involving the $A$ boson are dominated by its large couplings to $\tau$-leptons and top quarks, the $A$ boson coupling to down-type quarks is suppressed~\cite{Athron:2021iuf,PhysRevD.98.035001}.
This motivates the search for CP-odd Higgs bosons $A$ produced in proton--proton ($pp$) collisions at the LHC via gluon--gluon fusion involving a top-quark loop, with subsequent decay into a pair of $\tau$-leptons: $gg\to A\to \tau^+\tau^-$.
Within the flavour-aligned 2HDM, subject to the constraints described above, the cross-section for this production process scales with the square of the up-type quark coupling parameter $\zeta_\mathrm{u}$.
The current upper limit at \SI{95}{\%} confidence level (CL) on \azu is 0.5~\cite{PhysRevD.98.035001}.

In the analysis presented here, $A$ boson masses from \SI{20}{\giga\electronvolt} to \SI{90}{\giga\electronvolt} are considered. Previous searches for neutral Higgs bosons produced by gluon--gluon fusion in $pp$ collisions and with decays into $\tau$-lepton pairs were performed by the ATLAS and CMS collaborations and focused on $m_A$ ranges above a starting value of \SI{60}{\giga\electronvolt} to \SI{100}{\giga\electronvolt}~\cite{HIGG-2011-02,HIGG-2013-31,HDBS-2018-46,CMS-HIG-17-020,CMS-HIG-21-001}. The mass range from \SI{20}{\giga\electronvolt} to \SI{60}{\giga\electronvolt} is thus explored here for the first time for this production and decay mode.

\begin{figure}[!t]\centering
\includegraphics[width=0.7\textwidth]{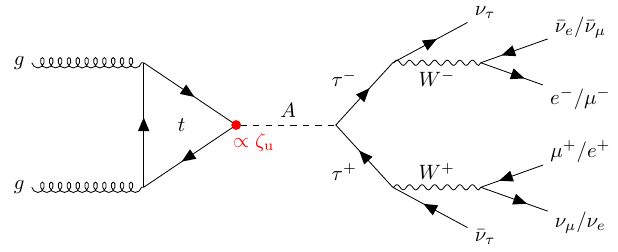}
\caption{Leading-order Feynman diagrams of the processes $gg\to A\to \tau^{+}\tau^{-} \to \mu^{+}\nu_{\mu} \bar{\nu}_{\tau} e^{-} \bar{\nu}_{e} \nu_{\tau}~,~e^{+}\nu_{e}\bar{\nu}_{\tau} \mu^-\bar{\nu}_{\mu}\nu_{\tau}$ within the flavour-aligned 2HDM with large $A$ boson couplings to $\tau$-leptons and top quarks, and suppressed couplings to down-type quarks.
The vertex in red describes the coupling of the $A$ boson to up-type quarks and scales with $\zeta_\mathrm{u}$.}
\label{fig:feynman}
\end{figure}

Since the $\tau$-leptons have transverse momentum values of $\pt\approx m_A/2$, the final-state hadrons, electrons, and muons from the $\tau$-lepton decays are in a transverse momentum range below $\pt\approx \SI{45}{\giga\electronvolt}$.
Because of the large background from QCD-induced processes in $pp$ collisions, the trigger $\pt$ thresholds for hadronically decaying $\tau$-leptons are typically well above these values. Electron--muon dilepton triggers, however, apply lower $\pt$ thresholds and are better suited for the light $A$ boson search.
Therefore, and also to suppress background from $Z\to e^{+}e^{-}$ and $Z\to \mu^{+}\mu^{-}$ decays, the analysis focuses on the fully leptonic signal process with exactly one electron and one muon in the final state, $gg\to A\to \tau^{+}\tau^{-} \to \mu^{+}\nu_{\mu} \bar{\nu}_{\tau} e^{-} \bar{\nu}_{e} \nu_{\tau}$ or $gg\to A\to \tau^{+}\tau^{-} \to e^{+}\nu_{e}\bar{\nu}_{\tau} \mu^-\bar{\nu}_{\mu}\nu_{\tau}$, as shown in \cref{fig:feynman}. These final states correspond to \SI{6.2}{\%} of all possible $\tau^{+}\tau^{-}$ final states~\cite{Workman:2022ynf}. The CP-odd nature of the $A$ boson and its decay kinematics are exploited to further reduce the background.


%
\section{ATLAS detector}
%
%

\newcommand{\AtlasCoordFootnote}{%
ATLAS uses a right-handed coordinate system with its origin at the nominal interaction point (IP)
in the centre of the detector and the \(z\)-axis along the beam pipe.
The \(x\)-axis points from the IP to the centre of the LHC ring,
and the \(y\)-axis points upwards.
Polar coordinates \((r,\phi)\) are used in the transverse plane,
\(\phi\) being the azimuthal angle around the \(z\)-axis.
The pseudorapidity is defined in terms of the polar angle \(\theta\) as \(\eta = -\ln \tan(\theta/2)\) and is equal to the rapidity
$ y = \frac{1}{2} \ln \left( \frac{E + p_z}{E - p_z} \right) $ in the relativistic limit.
Angular distance is measured in units of \(\Delta R \equiv \sqrt{(\Delta y)^{2} + (\Delta\phi)^{2}}\).}

The ATLAS detector~\cite{PERF-2007-01} at the LHC covers nearly the entire solid angle around the collision point.\footnote{\AtlasCoordFootnote}
It consists of an inner tracking detector surrounded by a thin superconducting solenoid, electromagnetic and hadronic calorimeters,
and a muon spectrometer incorporating three large superconducting air-core toroidal magnets.

The inner-detector system (ID) is immersed in a \qty{2}{\tesla} axial magnetic field
and provides charged-particle tracking in the range \(|\eta| < 2.5\).
The high-granularity silicon pixel detector covers the vertex region and typically provides four measurements per track,
the first hit generally being in the insertable B-layer (IBL) installed before Run~2~\cite{ATLAS-TDR-19,PIX-2018-001}.
It is followed by the SemiConductor Tracker (SCT), which usually provides eight measurements per track.
These silicon detectors are complemented by the transition radiation tracker (TRT),
which enables radially extended track reconstruction up to \(|\eta| = 2.0\).
The TRT also provides electron identification information
based on the fraction of hits (typically 30 in total) above a higher energy-deposit threshold corresponding to transition radiation.

The calorimeter system covers the pseudorapidity range \(|\eta| < 4.9\).
Within the region \(|\eta|< 3.2\), electromagnetic calorimetry is provided by barrel and
endcap high-granularity lead/liquid-argon (LAr) calorimeters,
with an additional thin LAr presampler covering \(|\eta| < 1.8\)
to correct for energy loss in material upstream of the calorimeters.
Hadronic calorimetry is provided by the steel/scintillator-tile calorimeter,
segmented into three barrel structures within \(|\eta| < 1.7\), and two copper/LAr hadronic endcap calorimeters.
The solid angle coverage is completed with forward copper/LAr and tungsten/LAr calorimeter modules
optimised for electromagnetic and hadronic energy measurements, respectively.

The muon spectrometer (MS) comprises separate trigger and
high-precision tracking chambers measuring the deflection of muons in a magnetic field generated by the superconducting air-core toroidal magnets.
The field integral of the toroids ranges between \num{2.0} and \qty{6.0}{\tesla\metre}
across most of the detector.
Three layers of precision chambers, each consisting of layers of monitored drift tubes, cover the region \(|\eta| < 2.7\),
complemented by cathode-strip chambers in the forward region, where the background is highest.
The muon trigger system covers the range \(|\eta| < 2.4\) with resistive-plate chambers in the barrel, and thin-gap chambers in the endcap regions.

The luminosity is measured mainly by the LUCID--2~\cite{LUCID2} detector that records Cherenkov light produced in the quartz windows of photomultipliers located close to the beampipe.

Events are selected by the first-level trigger system implemented in custom hardware,
followed by selections made by algorithms implemented in software in the high-level trigger~\cite{TRIG-2016-01}.
The first-level trigger accepts events from the \qty{40}{\MHz} bunch crossings at a rate below \qty{100}{\kHz},
which the high-level trigger further reduces in order to record complete events to disk at about \qty{1}{\kHz}.

A software suite~\cite{SOFT-2022-02} is used in data simulation, in the reconstruction
and analysis of real and simulated data, in detector operations, and in the trigger and data acquisition
systems of the experiment.


%
\section{Data and simulated event samples}
\label{sec:samples}
%

\subsection{Selection of data events}
The search presented in this paper uses the full
$pp$ collision dataset recorded by the ATLAS detector between 2015 and 2018 (\RunTwo),
with the LHC operating at a centre-of-mass energy of $\sqrt s = 13\,\mathrm{TeV}$ and a bunch spacing of $25\,\mathrm{ns}$.
In this period, the LHC delivered colliding beams with a peak instantaneous luminosity of $L=\SI{2.1e34}{cm^{-2}s^{-1}}$, achieved in 2018,
and an average number of $pp$ interactions per bunch crossing \muhat\ of 13, 25, 38, and 36 for 2015, 2016, 2017, and 2018 data, respectively~\cite{DAPR-2021-01}.
After applying beam, detector, and data-quality criteria
the total integrated luminosity of the data is $\SI{140}{\per\femto\barn}$.
The data were recorded using three partially overlapping electron--muon dilepton triggers \cite{TRIG-2018-05, TRIG-2018-01} that applied identification criteria based on shower shapes in the electromagnetic calorimeter.
The first trigger required lepton transverse momenta of $p_\mathrm{T}^e > 17\,\si{\giga\electronvolt}$ and $p_\mathrm{T}^\mu > 14\,\si{\giga\electronvolt}$,
while the other two triggers covered the lower $p_\mathrm{T}$ ranges, with either $p_\mathrm{T}^e > 7\,\si{\giga\electronvolt}$ and $p_\mathrm{T}^\mu > 24\,\si{\giga\electronvolt}$ or $p_\mathrm{T}^e > 24\,\si{\giga\electronvolt}$ and $p_\mathrm{T}^\mu > 8\,\si{\giga\electronvolt}$, respectively.
These trigger requirements have an efficiency of \SI{72}{\%} to \SI{92}{\%}, depending on the final state and the dataset, for
signal events passing the final signal-region selections.

\subsection{Monte Carlo event simulation}
\label{sec:MC_simulation}
All signal and background processes were modelled using Monte Carlo (MC) simulations, with the exception of backgrounds with fake or non-prompt leptons, which were estimated from data (see \cref{sec:fakebackground}).

Each sample was passed through the full \GEANT~\cite{Agostinelli:2002hh,SOFT-2010-01} simulation of the ATLAS detector and reconstructed with the same software as used for data. The only exceptions are the signal samples, for which the fast simulation ATLFAST-II~\cite{ATL-PHYS-PUB-2010-013} was used, where the \GEANT simulation of calorimeter showers is replaced by a parameterised simulation.
The gluon--gluon fusion signal-production process for this analysis, $gg\to A\to \tau^+\tau^-$, was simulated with \PYTHIA[8.2]~\cite{pythia8}, using leading-order (LO) matrix elements (ME) for $A$ bosons with masses of \SI{20}{\giga\electronvolt} to \SI{90}{\giga\electronvolt} in steps of \SI{10}{\giga\electronvolt}, as well as for \SI{25}{\giga\electronvolt}, \SI{75}{\giga\electronvolt} and \SI{85}{\giga\electronvolt}. Both $\tau$-leptons were forced to decay leptonically.
The parton shower, hadronisation, and underlying event were modelled with parameter values set according to the A14 tune and with the \NNPDF[2.3lo] set of parton distribution functions (PDF)~\cite{Ball:2012cx}. The signal's production cross-section was calculated at $\mathrm{N^3LO}$ with \texttt{ggHiggs}~\cite{Ball_2013,Bonvini_2014,Bonvini_2016,Ahmed_2016,Bonvini_2018,Bonvini_2013,Harlander_2002,Harlander:2009mq,Harlander:2009my,anastasiou2014,Anastasiou:2016cez,Mistlberger_2018} version~4.0, setting the coupling parameter of the $A$ boson to up-type quarks to $\cot\beta=\zeta_\mathrm{u}=1$, where $\cot\beta$ is the ratio of the vacuum expectation values of the two Higgs doublets and is used by \texttt{ggHiggs} to parameterise the fermionic coupling parameter.
The cross-section values are shown in \cref{tab:signal_xsec_unc_eff} for the relevant mass range.

\begin{table}[!t]
\caption{$\mathrm{N^3LO}$ prediction of
the cross-section for $A$ boson production, $\sigma (pp \to A + X)$, via gluon--gluon fusion in the relevant mass range.
Values were calculated with \texttt{ggHiggs}~\cite{Ball_2013,Bonvini_2014,Bonvini_2016,Ahmed_2016,Bonvini_2018,Bonvini_2013,Harlander_2002,Harlander:2009mq,Harlander:2009my,anastasiou2014,Anastasiou:2016cez,Mistlberger_2018} version~4.0,
setting the fermionic coupling strength parameters of the $A$ boson to $\cot\beta=\zeta_\mathrm{u}=\zeta_\ell=1$.
The product of the signal acceptance $\mathcal{A}_\mathrm{signal}$ and efficiency $\varepsilon_\mathrm{signal}$ is the fraction of events
generated for gluon--gluon fusion production of the CP-odd Higgs boson $A$ (which is forced to decay into a leptonically decaying $\tau^+\tau^-$
pair) that are reconstructed in the detector and accepted by the analysis event selection described in \cref{sec:eventselection}.
}  \centering
\begin{tabular}{c|cccccccc}
\hline\hline
\rule{0pt}{14pt} $m_A$ [\si{\giga\electronvolt}] & \multicolumn{1}{c}{20} & \multicolumn{1}{c}{25} & \multicolumn{1}{c}{30} & \multicolumn{1}{c}{40} & \multicolumn{1}{c}{50} & \multicolumn{1}{c}{60} & \multicolumn{1}{c}{70} & \multicolumn{1}{c}{75}\\\hline\hline
\rule{0pt}{14pt} $\sigma (pp \to A + X)$ [\si{\pico\barn}] & 1854 & 1394 & 1090 & ~722 & ~514 & ~385 & ~299 & 266 \\\hline
\rule{0pt}{14pt} $\mathcal{A}_\mathrm{signal} \times \varepsilon_\mathrm{signal}$  [\si{\percent}] & 0.017 & 0.038 & 0.060 & 0.094 & 0.10  & 0.091 & 0.077 & 0.20 \\\hline\hline
\multicolumn{8}{l}{\color{white}.}
\end{tabular}
\begin{tabular}{c|ccc}
\hline\hline
\rule{0pt}{14pt} $m_A$ [\si{\giga\electronvolt}] &  \multicolumn{1}{c}{80} & \multicolumn{1}{c}{85} & \multicolumn{1}{c}{90} \\\hline\hline
\rule{0pt}{14pt} $\sigma (pp \to A + X)$ [\si{\pico\barn}] & ~239 & 215 & ~195\\\hline
\rule{0pt}{14pt} $\mathcal{A}_\mathrm{signal} \times \varepsilon_\mathrm{signal}$ [\si{\percent}] & 0.20 & 0.19 & 0.18 \\\hline\hline
\multicolumn{3}{l}{\color{white}.}
\end{tabular}

\label{tab:signal_xsec_unc_eff}
\end{table}

The MC-simulated background processes considered in this analyses include those with vector bosons (diboson and $V$+jets, where $V=W,Z$), top quarks (\ttbar,  $s$- and $t$-channel single-top, and associated production of a top quark and a $W$ boson ($tW$)), and SM Higgs bosons, which are all modelled differently.

The $V+$jets processes were simulated with the \SHERPA[2.2.1]~\cite{Bothmann:2019yzt} generator, with next-to-leading-order (NLO) MEs for up to two partons and
LO MEs for up to four partons. The diboson background processes were simulated with NLO MEs for up to one parton and
LO MEs for up to three partons, using the \SHERPA[2.2.1]~\cite{Bothmann:2019yzt} or \SHERPA[2.2.2]~\cite{Bothmann:2019yzt} generator, depending on the process.
The top-quark-related backgrounds were modelled using the \POWHEGBOX[v2]~\cite{Frixione:2007nw,Nason:2004rx,Frixione:2007vw,Alioli:2010xd}
generator at NLO accuracy in QCD with the \NNPDF[3.0nlo] PDF set~\cite{Ball:2014uwa}. Both the $tW$ and $s$-channel single-top events were generated using the five-flavour scheme. However, a four-flavour scheme was used for  $t$-channel single-top events. For the \ttbar sample, a value of $1.5\,m_t$ was assigned to the resummation damping factor \hdamp, which controls the matching of the \POWHEG ME to the parton shower and thus effectively regulates the high-\pt radiation against which the \ttbar system recoils.

In order to model the parton shower, hadronisation, and underlying event,
the top-quark background events were interfaced to \PYTHIA[8.230] with parameter values set according to the A14 tune~\cite{ATL-PHYS-PUB-2014-021}, while the \SHERPA parton shower~\cite{Schumann:2007mg}, using the \MEPSatNLO prescription~\cite{Hoeche:2011fd,Hoeche:2012yf,Catani:2001cc,Hoeche:2009rj}
and the set of tuned parameters developed by the \SHERPA authors, was applied to $W/Z$+jets and diboson background events.
The \NNPDF[2.3lo] PDF set was used for both types of background.

SM Higgs boson production was simulated using \POWHEGBOX[v2]~\cite{Hamilton:2013fea,Hamilton:2015nsa,Alioli:2010xd,Nason:2004rx,Frixione:2007vw}, interfaced with \PYTHIA[8] for parton showering and non-perturbative effects. The \PDFforLHC[15nlo] PDF set~\cite{Butterworth:2015oua} and the \AZNLO tune~\cite{STDM-2012-23} of \PYTHIA[8] were used. SM Higgs boson production via gluon--gluon fusion and via vector-boson fusion was simulated at next-to-next-to-leading-order (NNLO) and NLO accuracy in QCD, respectively. The \POWHEG prediction is accurate to NLO for $Vh$ boson plus one-jet production.
The loop-induced $gg\to Zh$ process was generated separately at LO. The normalisation of all SM Higgs boson samples accounts for the decay branching ratio calculated with HDECAY~\cite{Djouadi:1997yw,Spira:1997dg,Djouadi:2006bz} and \PROPHECY~\cite{Bredenstein:2006ha,Bredenstein:2006rh,Bredenstein:2006nk-fixed}.
The MC prediction for SM Higgs boson production via gluon--gluon fusion was normalised to the cross-section calculated to $\mathrm{N^3LO}$ in QCD
with NLO electroweak (EW) corrections~\cite{deFlorian:2016spz,Anastasiou:2016cez,Anastasiou:2015ema,Dulat:2018rbf,Harlander:2009mq,Harlander:2009bw,Harlander:2009my,Pak:2009dg,Actis:2008ug,Actis:2008ts,Bonetti:2018ukf},
while that for production via vector-boson fusion was normalised to the cross-section calculated to approximate-NNLO in QCD with NLO EW corrections~\cite{Ciccolini:2007jr,Ciccolini:2007ec,Bolzoni:2010xr}.
The MC predictions for $q\bar{q}/qg \to Vh$ and $gg \to Zh$ were normalised to the cross-section calculated to NNLO in QCD with NLO EW corrections or to NLO and next-to-leading-logarithm accuracy in QCD, respectively~\cite{Ciccolini:2003jy,Brein:2003wg,Brein:2011vx,Altenkamp:2012sx,Denner:2014cla,Brein:2012ne,Harlander:2014wda}.

To model additional $pp$ interactions in the same or neighbouring bunch crossings (pile-up), the hard-scattering events were overlaid with a set of minimum-bias interactions generated using \PYTHIA[8.186]~\cite{Sjostrand:2007gs} and the \NNPDF[2.3lo] PDF set with the A3 set of tuned parameters~\cite{ATL-PHYS-PUB-2016-017}.
Finally, all simulated events are reweighted to match the instantaneous luminosity spectrum in data.


%
%

%
\section{Event reconstruction}
%
%

%
Charged-particle tracks measured in the ID are used to reconstruct interaction vertices~\cite{ATL-PHYS-PUB-2015-026}, and the one with the highest sum of squared transverse momenta $\sum \pT^{2}$ of the associated tracks is selected as the primary vertex of the hard interaction.

Electrons are reconstructed from topological clusters of energy deposits in the electromagnetic calorimeter which are matched to a track reconstructed in the ID and are required to satisfy \textit{Loose} identification and \textit{Loose} isolation criteria~\cite{EGAM-2018-01}, based on track and shower-shape discriminants.

Muons are reconstructed from signals in the muon spectrometer matched to tracks inside the ID. They are also required to satisfy \textit{Loose} identification and \textit{Loose} isolation criteria based on track information~\cite{MUON-2018-03}.

Jets are reconstructed from particle-flow objects~\cite{PERF-2015-09} using the anti-$k_t$ algorithm~\cite{Cacciari:2008gp} with radius parameter $R=0.4$.
This algorithm uses  noise-suppressed positive-energy topological clusters in the calorimeter after subtracting energy deposits associated with primary-vertex-matched tracks and including the momenta of those tracks in the clustering instead, thereby improving the jet energy measurement. The jet energy scale is calibrated to the particle level using simulation and further corrected with in-situ methods~\cite{JETM-2018-05}. Cleaning criteria are used to identify jets arising from non-collision backgrounds or noise in the calorimeters~\cite{ATLAS-CONF-2015-029}. A dedicated jet-vertex-tagger algorithm~\cite{PERF-2014-03} is used to remove jets within $|\eta| < 2.5$ that are identified as not being associated with the primary vertex of the hard interaction. Similarly, a dedicated algorithm is used to suppress such \enquote{pile-up} jets, arising from additional softer $pp$ interactions, in the forward region~\cite{PERF-2016-06}.

Jets containing $b$-hadrons are identified using the DL1r~\cite{FTAG-2019-07} \textit{b}-tagging algorithm. A jet is classified as a \textit{b}-tagged jet if the DL1r score is above a certain threshold, referred to as an \enquote{operating point}. In this analysis, the operating point is chosen to obtain an average expected efficiency of \SI{85}{\%} for \textit{b}-jet identification, as determined in simulated \ttt events.

Since the different physical objects are reconstructed independently, multiple objects may be reconstructed from the same detector signal,
whereas a signal rarely has contributions from multiple objects.
To achieve orthogonality, an overlap removal procedure is performed on objects with small angular separation $\Delta R$ to remove the
object with lower signal purity.
Muons are favoured over electrons if one lies within a $\Delta R = 0.2$ cone around the other, and leptons are favoured over jets if their
angular separation is $\Delta R < 0.4$.

The missing transverse momentum $\vec{p}_\text{T}^{\,\text{miss}}$ (with magnitude \met) is reconstructed as the negative vector sum of the transverse momenta of leptons, visible hadronic $\tau$-lepton decay products (\tauhadvis)~\cite{ATL-PHYS-PUB-2015-045}, jets, and a \enquote{soft term}. The \enquote{soft term} is calculated as the vector sum of the \pT of tracks matched to the primary vertex but not to reconstructed leptons, \tauhadvis candidates, or jets~\cite{JETM-2020-03}.

An advanced likelihood-based algorithm named the Missing Mass Calculator (MMC)~\cite{mmc,HIGG-2019-09} is used to estimate the invariant mass, $m_\mathrm{MMC}$, of the $\tau$-lepton pair under the assumption that only neutrinos originating from the $\tau$-lepton decays contribute to \met.
In order to resolve the remaining ambiguities in the reconstruction of the neutrino momenta, the method incorporates additional knowledge of the decay kinematics in terms of probability density functions, such as distributions of the expected angular distances between the decay products of the $\tau$-leptons.


%
%
%

%

\section{Signal selection}
\label{sec:signalselection}
%

\label{sec:eventselection}
This analysis focuses on final states with exactly one electron and one muon with opposite charges and $p_\mathrm{T}$ greater than \SI{10}{\giga\electronvolt}.
Events are required to have been accepted by at least one of the electron--muon triggers that were active during the data-taking periods (see \cref{sec:samples}).
The leptons firing the trigger must match the two reconstructed leptons with the highest \pt in the event.
The matching is performed according to the angular distance $\Delta R$ between the trigger signature and the reconstructed lepton, requiring $\Delta R<0.07$ for electrons and $\Delta R<0.1$ for muons.
Furthermore, the selected electron or muon candidates are required to satisfy \textit{Medium} identification and \textit{Tight} or \textit{Tight\_VarRad} isolation criteria~\cite{EGAM-2018-01,MUON-2018-03}, respectively, in order to increase the signal region's purity in true reconstructed leptons. Additionally, events with leptons satisfying the \textit{Loose} criteria are used in the fake-lepton estimation.

The signal region (SR) definition depends on the probed $A$ mass value: masses from \SI{20}{} to \SI{75}{\giga\electronvolt} are examined in the low-mass SR, and masses from \SI{75}{} to \SI{90}{\giga\electronvolt} are examined in the high-mass SR. A common requirement for both SRs is that the number of $b$-tagged jets, $n_{b\text{-jets}}$, must be zero. This significantly reduces the background from top quarks. In order to optimise the signal sensitivity, the low-mass SR and the high-mass SR have individualised selection criteria for \met, the total transverse mass defined as $m_\text{T}^\text{tot}=\sqrt{\left(p_\text{T}^e+p_\text{T}^\mu+E_\text{T}^\text{miss}\right)^2-\left(\vec{p}_\text{T}^{\,e}+\vec{p}_\text{T}^{\,\mu}+\vec{p}_\text{T}^{\,\text{miss}}\right)^2}$, the angular distance between the charged leptons \drll, and $m_\mathrm{MMC}$.
The selection criteria are summarised in \cref{tab:srcuts}.
The \mtt criterion primarily removes diboson events and events with top quarks.
The requirement on \drll reduces the \ztt background, motivated by the mass, spin and CP properties of the $A$ boson signal, and requires the emitted leptons to have a small angular separation.
This kinematic configuration also enhances the fraction of $A$ bosons produced with sizeable transverse momentum $p_\mathrm{T}^A$.
Thus, the $p_\mathrm{T}^A$ distribution starts at approximately $100$ to \SI{200}{\giga\electronvolt}, depending on the mass hypothesis.
Therefore, hadronic jets are emitted in the opposite direction to conserve transverse momentum, meaning the process $pp \rightarrow A + \text{jets}$ is preferred for this topology (considering only the gluon--gluon fusion production mode as suggested by the flavour-aligned 2HDM).

\begin{table}[!t]
\caption[Overview of the requirements for the most important analysis regions]{Overview of the selection criteria for the most important regions: the low-mass and high-mass signal regions (SR), the top control region (TCR), the \ztt control region (ZCR) and the fake-lepton validation region (FVR). For the final results of the analysis, \textit{Medium} identification and \textit{Tight} isolation criteria are used~\cite{EGAM-2018-01,MUON-2018-03}. For the fake-lepton background estimation, \textit{Loose} criteria are used additionally.}
\footnotesize
\makebox[1 \textwidth][c]{
\begin{tabular}{c|cc|cc|c}
\hline\hline
\rule{0pt}{12pt} & \multicolumn{2}{c|}{SR}\centering& TCR & ZCR & FVR\\
\rule{0pt}{12pt} & \centering{Low-mass} & {High-mass}  & & &\\
\hline\hline
\rule{0pt}{12pt} $|\eta_e|$  &  \multicolumn{2}{c|}{$[0,1.37]\cup [1.52,2.47]$} & \multicolumn{2}{c|}{$[0,1.37]\cup [1.52,2.47]$} & $[0,1.37]\cup [1.52,2.47]$ \\
\rule{0pt}{12pt} $|\eta_\mu|$ & \multicolumn{2}{c|}{$[0,2.7]$} &  \multicolumn{2}{c|}{$[0,2.7]$} & $[0,2.7]$ \\
\rule{0pt}{12pt} $|\eta_j|$  &  \multicolumn{2}{c|}{$[0,4.5]$} &  \multicolumn{2}{c|}{$[0,4.5]$} & $[0,4.5]$ \\
\rule{0pt}{12pt} $\pt^\ell$  & \multicolumn{2}{c|}{{$>\SI{10}{\giga\electronvolt}$}} & \multicolumn{2}{c|}{{$>\SI{10}{\giga\electronvolt}$}} & $>\SI{10}{\giga\electronvolt}$ \\
\rule{0pt}{12pt} $\pt^j$  & \multicolumn{2}{c|}{{$>\SI{20}{\giga\electronvolt}$}} & \multicolumn{2}{c|}{{$>\SI{20}{\giga\electronvolt}$}} & $>\SI{20}{\giga\electronvolt}$ \\
\rule{0pt}{12pt} \met  & ~{$>\SI{50}{\giga\electronvolt}$} & ~{$>\SI{30}{\giga\electronvolt}$}& ~{$>\SI{30}{\giga\electronvolt}$} & $-$ & $-$\\
\rule{0pt}{12pt} $m_\text{T}^\text{tot}$ & ~{$<\SI{45}{\giga\electronvolt}$} & ~{$<\SI{65}{\giga\electronvolt}$} & ~{$<\SI{65}{\giga\electronvolt}$}  & ~{$<\SI{65}{\giga\electronvolt}$}& ~{$<\SI{65}{\giga\electronvolt}$}\\
\rule{0pt}{12pt} \drll  & {$<\num{0.7}$} & {$<\num{1.0}$} & {$<\num{1.0}$} & {$>\num{1.4}$}& {$>\num{1.4}$}\\
\rule{0pt}{12pt} $m_\text{MMC}$  & {$>\SI{0}{\giga\electronvolt}$} & {$>\SI{35}{\giga\electronvolt}$\&} & {$>\SI{0}{\giga\electronvolt}$} & {$>\SI{0}{\giga\electronvolt}$}& {$>\SI{0}{\giga\electronvolt}$}\\
\rule{0pt}{8pt} &  & {$<\SI{130}{\giga\electronvolt}$} & & &\\
\rule{0pt}{12pt} $q_e\times q_\mu$  & {$-1$} & {$-1$}  & {$-1$} & {$-1$}& {$1$}\\
\rule{0pt}{12pt} $n_{b\text{-jets}}$  & \num{0} & \num{0}  & {$\geq2$} & \num{0}& \num{0}\\
\hline\hline
\end{tabular}
}
\label{tab:srcuts}
\end{table}


%
\newpage
\section{Background estimation}
%
%

There are two primary sources of background events associated with the event selection described in \cref{sec:signalselection}: events with electrons and muons produced in $W$, $Z/\gamma^*$, SM Higgs boson or top-quark decays (prompt leptons) and events that have at least one lepton arising from another source (fake and non-prompt leptons).
In the latter case, electrons and muons originate from improper reconstruction of other physics objects such as hadronic jets,
heavy-flavour hadron decays or photon conversions.
All processes associated with the prompt-lepton background are predominantly well modelled by simulation and therefore estimated from MC event samples, described in \cref{sec:MC_simulation}.
Two control regions, enriched in top-quark (TCR) and \ztt (ZCR) events, are designed to control and validate the major backgrounds modelled by MC simulation.
The fake-lepton validation region (FVR) is designed and used to estimate and validate the fake/non-prompt-lepton background using a data-driven method. A summary of the requirements for all regions is given in \cref{tab:srcuts}.

\subsection{Background with prompt leptons}
\label{sec:promptbackground}
The ZCR is used to control the main background in this analysis, which originates from \ztt decays. The requirements for this region are to have an abundance of \ztt decays, to be orthogonal to the SR, and to have low contamination from signal events while being kinematically close to the SR. This is achieved by having selection criteria similar to those for the SR, but requiring $\drll>1.4$ because the two leptons originating from the $Z$ boson are well separated. Additionally, the \met criterion is dropped in order to obtain a large sample of \ztt decays with low signal contamination.

The distributions of some variables in the ZCR showed mismodelling of \ztt events.
In particular, the event count with the number of jets, $n_\mathrm{jets}$, equal to 0 is underestimated in simulation and increasingly overestimated for $n_\mathrm{jets} > 1$.
Also, for high values of the transverse momentum of the ditau system,\footnote{The transverse momentum of the ditau system is defined as the magnitude of the vector sum of the transverse momenta of the electron and the muon and of the missing transverse momentum: $p_\text{T}^{\tau\tau}=|\vec{p}_\text{T}^{\,e}+\vec{p}_\text{T}^{\,\mu}+\vec{p}_\text{T}^{\,\text{miss}}|$} $p_\mathrm{T}^{\tau\tau}>50\,\si{\giga\electronvolt}$, and for \met between \num{50} and \SI{100}{\giga\electronvolt} a surplus of background events is observed.
This region is especially important for this analysis because the high-mass SR (low-mass SR) requires $\met > 30\,(50)\,\si{\giga\electronvolt}$, and \ztt events are the main background in the SR. Therefore, a data-driven reweighting method, similar to the one used in Ref.~\cite{STDM-2017-05}, is chosen to obtain better agreement between the prediction and data in the ZCR.
The weights are calculated as the ratio of the expected number of \ztt events to the simulated number in the ZCR and binned in the number of jets. The expected number of events is obtained from data by subtracting the predicted number of other simulated background events, which do not originate from \ztt. The fake-lepton background (see \cref{sec:fakebackground}) must also be taken into account, so the efficiencies for the \enquote{matrix method} described in the next section must be determined prior to the calculation of the $Z\to\tau^+\tau^-$ weights. The weights range from $0.8$ to $1.12$ with a tendency to be closer to unity for a smaller number of jets.
The reweighting not only resolves the mismodeling of $n_\mathrm{jets}$ but also improves the data-to-background agreement of the $p_\mathrm{T}^{\tau\tau}$ and \met distributions in the ZCR.

\begin{figure}[!t] \centering
\subfloat[]{\includegraphics[width=0.49\textwidth]{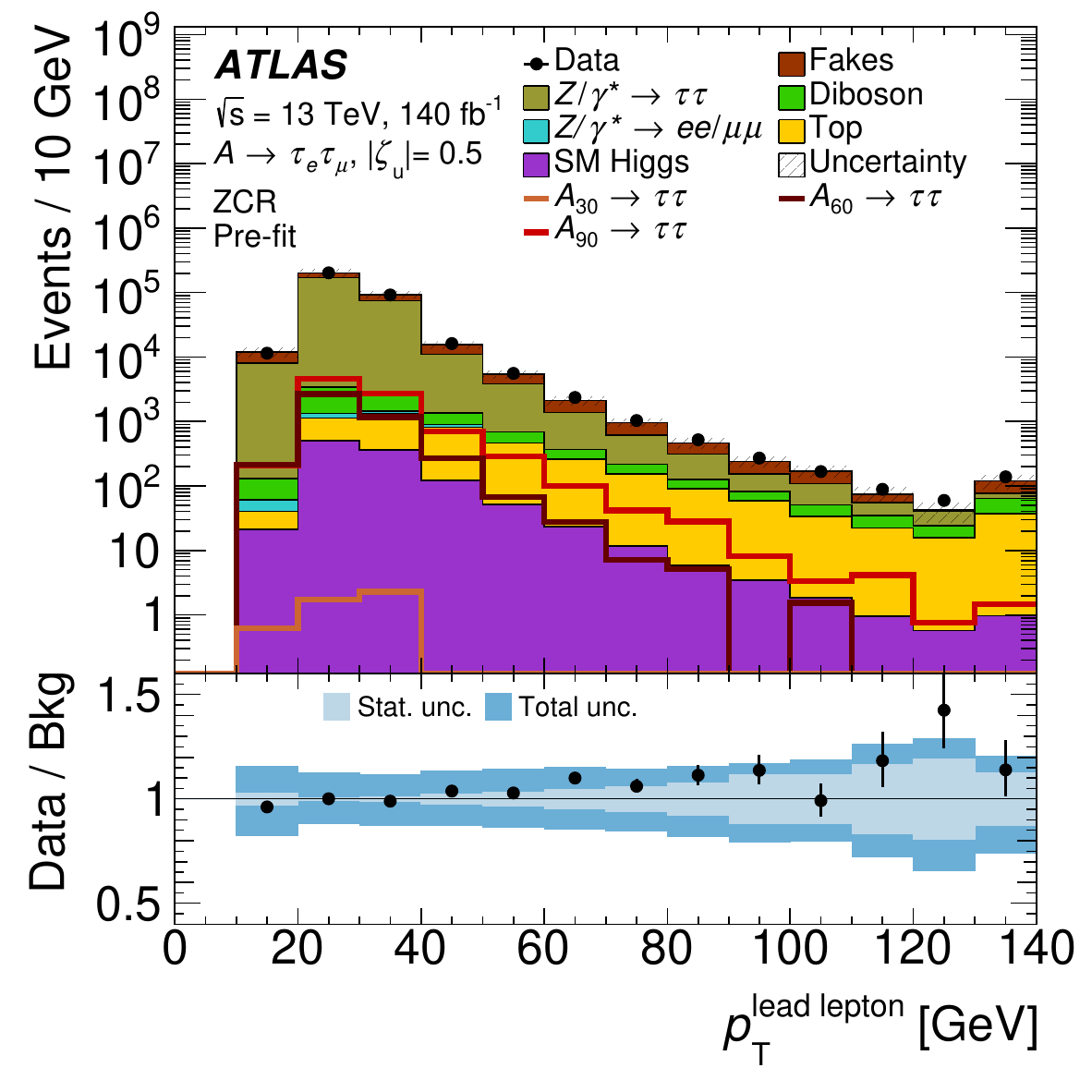}
}
\subfloat[]{\includegraphics[width=0.49\textwidth]{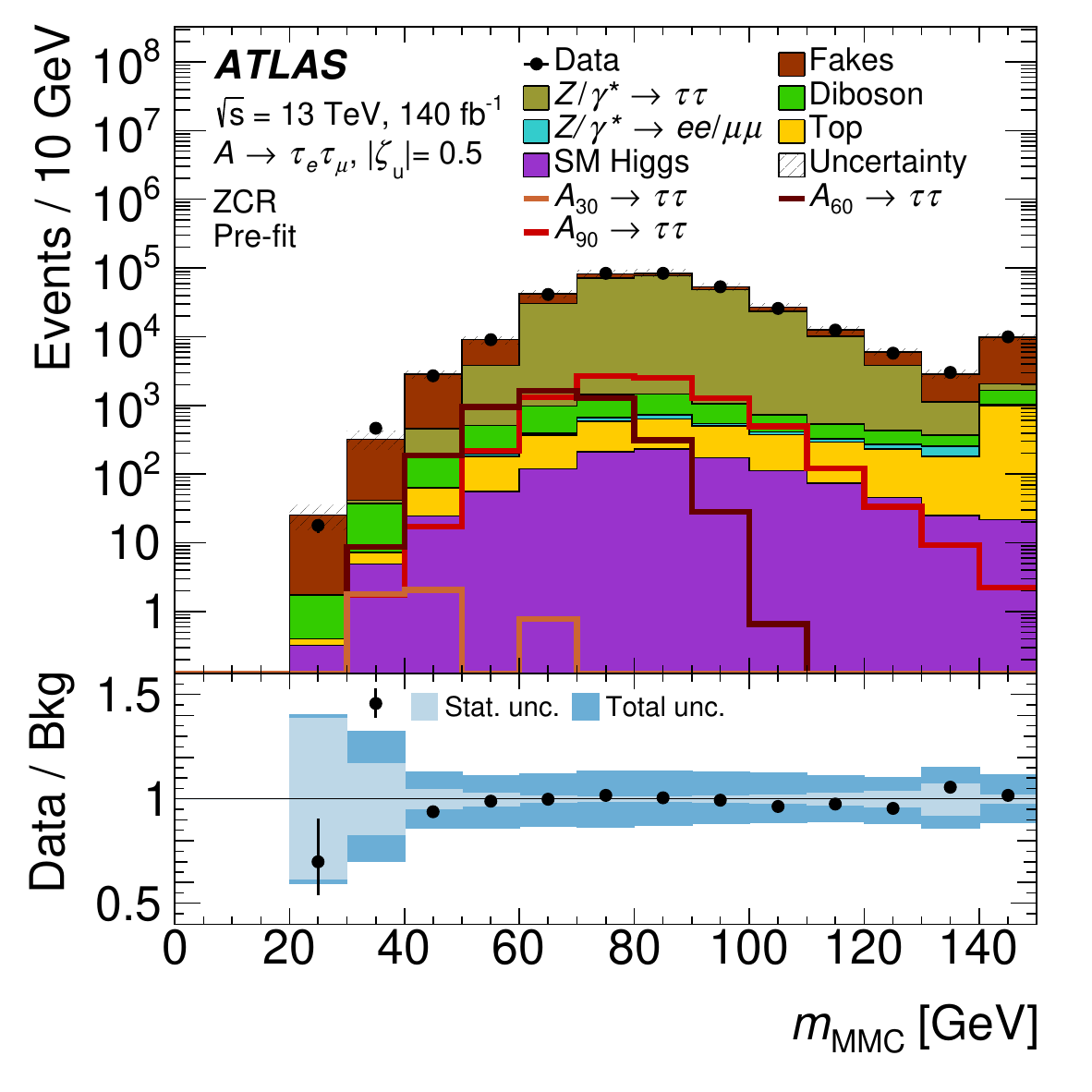}
}
\caption{Pre-fit variable distributions in the ZCR after applying all
selection criteria: (a) leading-lepton \pt and (b) \mmc mass estimator.
\commonselectioncaptionFigureCR{}}
\label{fig:ZVR-selection-1}
\end{figure}

\begin{figure}[!t] \centering
\subfloat[]{\includegraphics[width=0.49\textwidth]{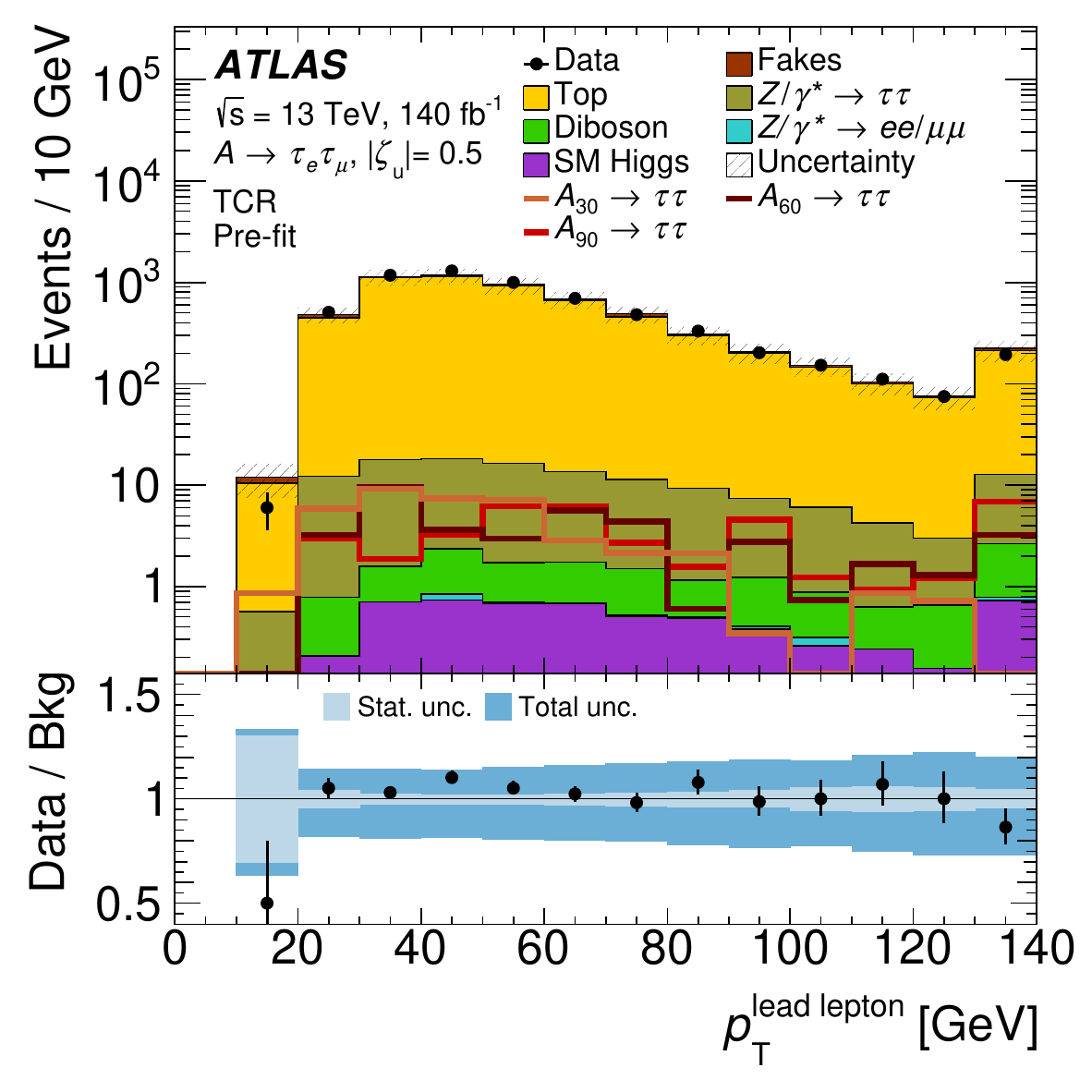}
}
\subfloat[]{\includegraphics[width=0.49\textwidth]{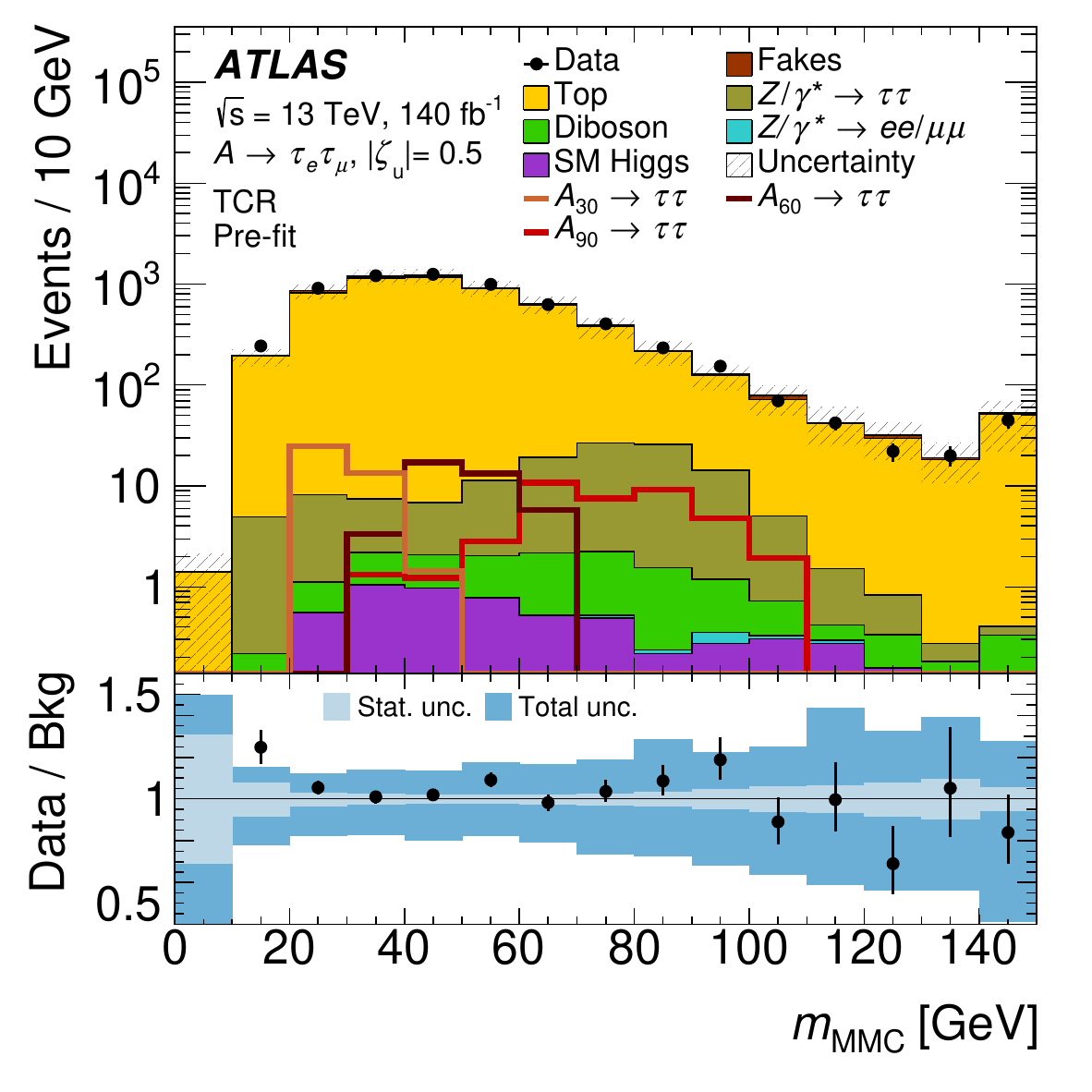}
}
\caption{Pre-fit variable distributions in the TCR after applying all
selection criteria: (a) leading-lepton \pt and (b) \mmc mass estimator.
\commonselectioncaptionFigureCR{}}
\label{fig:TVR-selection-1}
\end{figure}

The TCR is constructed to control the background resulting from processes involving top quarks, which produce one of the largest backgrounds in this analysis. It has almost the same requirements as the high-mass SR, but at least two $b$-tagged jets are required. Requiring $n_{b\text{-jets}}{\geq}1$ was also tested, but the contamination from signal events was too large. Furthermore, a small improvement to the modelling is incorporated by reweighting each \ttt sample so that its top-quark \pt distribution matches that calculated to NNLO in QCD with NLO EW corrections~\cite{Czakon_2017}.
\cref{fig:ZVR-selection-1} and \cref{fig:TVR-selection-1} show the validation of \Zjets and top-quark backgrounds, respectively. In both cases, the observed data agree well with the estimated background within their respective uncertainties.

\subsection{Background with non-prompt leptons}
\label{sec:fakebackground}
The FVR is constructed to perform and validate the fake-lepton estimation in a region enriched in non-prompt leptons.
It has the same requirements as the ZCR, except that the two leptons are required to have the same charge.
This region can be used for a closure test of the fake-lepton estimation, and systematic uncertainties can be extracted.

To estimate the total contribution of events with fake or non-prompt leptons, as well as leptons with misidentified charge (hereinafter collectively referred to as fake-lepton background), a data-driven matrix method~\cite{EGAM-2019-01} is used, which is especially useful in cases where events with either zero, one, or two fake leptons are expected.
Two categories of events are selected, satisfying baseline (\textit{Loose} isolation and \textit{Loose} identification) and tight (\textit{Tight} isolation and \textit{Medium} identification) lepton selection requirements as described in \cref{sec:eventselection}.
The efficiencies for fake leptons, $\varepsilon^{\ell}_\mathrm{fake}$, and for real leptons, $\varepsilon^{\ell}_\mathrm{real}$, that meet the tight criteria are used to estimate the rate of events with fake leptons from the numbers of tight and non-tight events passing the baseline selection.
This is done by inverting the equation
\begin{align*}
\left ( \begin{matrix} N^\mathrm{TT}_\mathrm{XX} \\[0.4em] N^\mathrm{T\overline{T}}_\mathrm{XX} \\[0.4em] N^\mathrm{\overline{T}T}_\mathrm{XX}  \\[0.4em] N^\mathrm{\overline{T} \overline{T}}_\mathrm{XX} \end{matrix} \right )
&=
\left ( \begin{matrix} \varepsilon^e_\mathrm{real} \varepsilon^\mu_\mathrm{real} & \,\varepsilon^e_\mathrm{real} \varepsilon^\mu_\mathrm{fake} & \,\varepsilon^e_\mathrm{fake} \varepsilon^\mu_\mathrm{real} & \,\varepsilon^e_\mathrm{fake} \varepsilon^\mu_\mathrm{fake} \\[0.4em] \varepsilon^e_\mathrm{real} \bar \varepsilon^\mu_\mathrm{real} & \,\varepsilon^e_\mathrm{real} \bar \varepsilon^\mu_\mathrm{fake} & \,\varepsilon^e_\mathrm{fake} \bar \varepsilon^\mu_\mathrm{real} & \,\varepsilon^e_\mathrm{fake} \bar \varepsilon^\mu_\mathrm{fake} \\[0.4em] \bar \varepsilon^e_\mathrm{real} \varepsilon^\mu_\mathrm{real} & \,\bar \varepsilon^e_\mathrm{real} \varepsilon^\mu_\mathrm{fake} & \,\bar \varepsilon^e_\mathrm{fake} \varepsilon^\mu_\mathrm{real} & \,\bar \varepsilon^e_\mathrm{fake} \varepsilon^\mu_\mathrm{fake} \\[0.4em] \bar \varepsilon^e_\mathrm{real} \bar \varepsilon^\mu_\mathrm{real} & \,\bar \varepsilon^e_\mathrm{real} \bar \varepsilon^\mu_\mathrm{fake} & \,\bar \varepsilon^e_\mathrm{fake} \bar \varepsilon^\mu_\mathrm{real} & \,\bar \varepsilon^e_\mathrm{fake} \bar \varepsilon^\mu_\mathrm{fake}  \end{matrix} \right )
\cdot{}
\left ( \begin{matrix} N^\mathrm{LL}_\mathrm{RR} \\[0.4em] N^\mathrm{LL}_\mathrm{RF} \\[0.4em] N^\mathrm{LL}_\mathrm{FR}  \\[0.4em] N^\mathrm{LL}_\mathrm{FF} \end{matrix} \right ),
\end{align*}
where $\bar \varepsilon^{\ell}_\mathrm{real/fake} = 1 - \varepsilon^{\ell}_\mathrm{real/fake}$.
The indices of the event yields $N$ are to be read as follows.
The upper indices describe the tightness of the leptons, where T means that the lepton is required to be tight, $\mathrm{\overline{T}}$ that it has to be non-tight, and L that the lepton is part of the baseline selection without further requirements on the tightness.
The lower indices indicate whether an event comes from data or from MC simulation, and in the latter case whether the lepton is real (R), fake (F), or without a requirement on a certain \enquote{truth} state (X) in the MC event record.
The first index always describes the electron, the second one the muon.
For example, $N^\mathrm{{L}{T}}_\mathrm{RX}$ is the number of MC events with a real electron that passes the baseline selection and a tight muon, and $N^\mathrm{{T}{\overline{T}}}_\mathrm{Data}$ is the number of data events with a tight electron and a non-tight muon. The total fake-lepton background is then the sum of the fake-electron, fake-muon, and double-fake-lepton backgrounds
\begin{align*}
N_\mathrm{fake}^\mathrm{TT} =  \varepsilon^{e}_\mathrm{real} \varepsilon^{\mu}_\mathrm{fake} N^\mathrm{LL}_\mathrm{RF} + \varepsilon^{e}_\mathrm{fake} \varepsilon^{\mu}_\mathrm{real} N^\mathrm{LL}_\mathrm{FR} + \varepsilon^{e}_\mathrm{fake} \varepsilon^{\mu}_\mathrm{fake} N^\mathrm{LL}_\mathrm{FF}.
\end{align*}

\begin{figure}[!t]\centering
\subfloat[]{\includegraphics[width=0.46\linewidth]{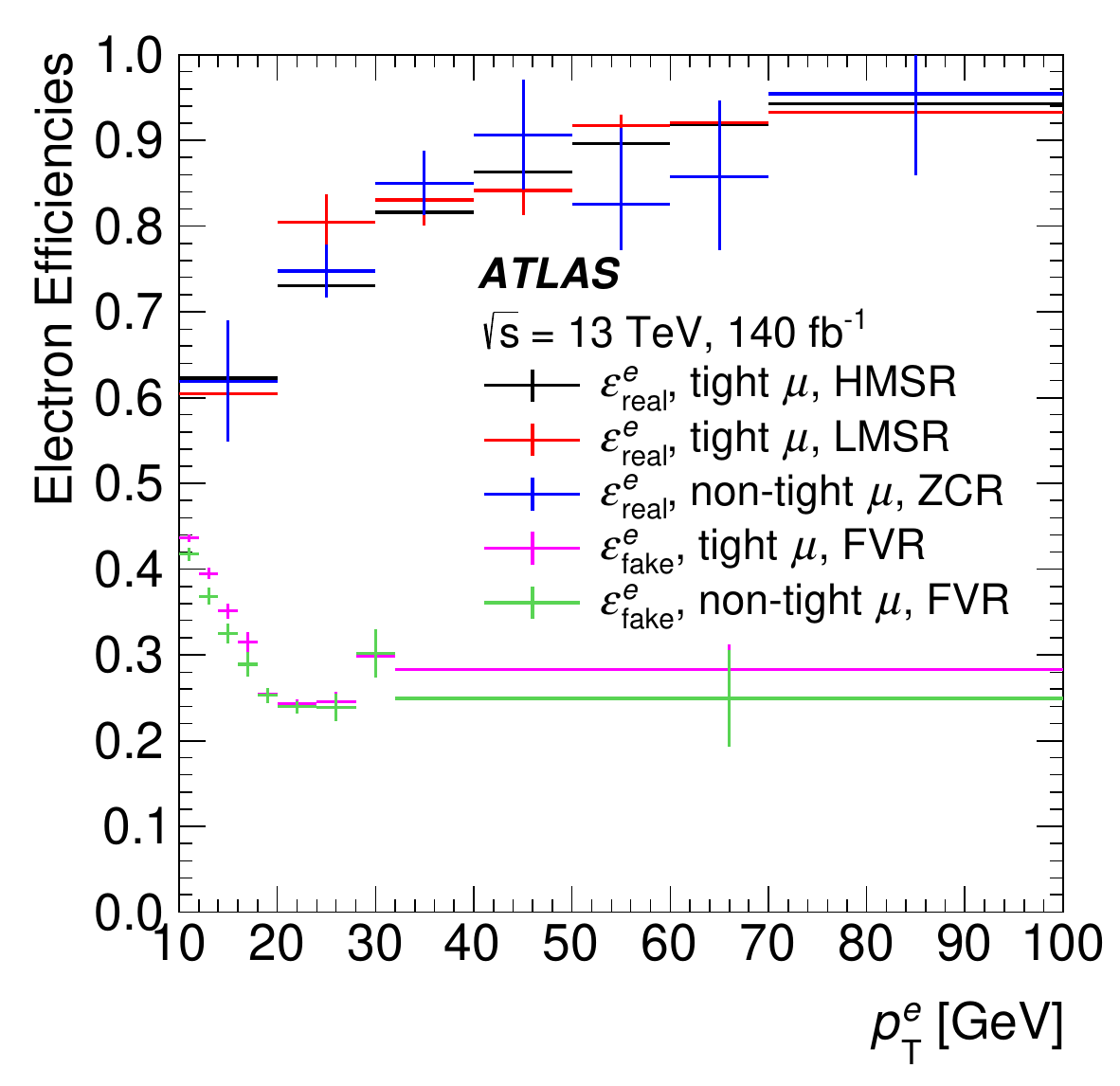}
}
\subfloat[]{\includegraphics[width=0.46\linewidth]{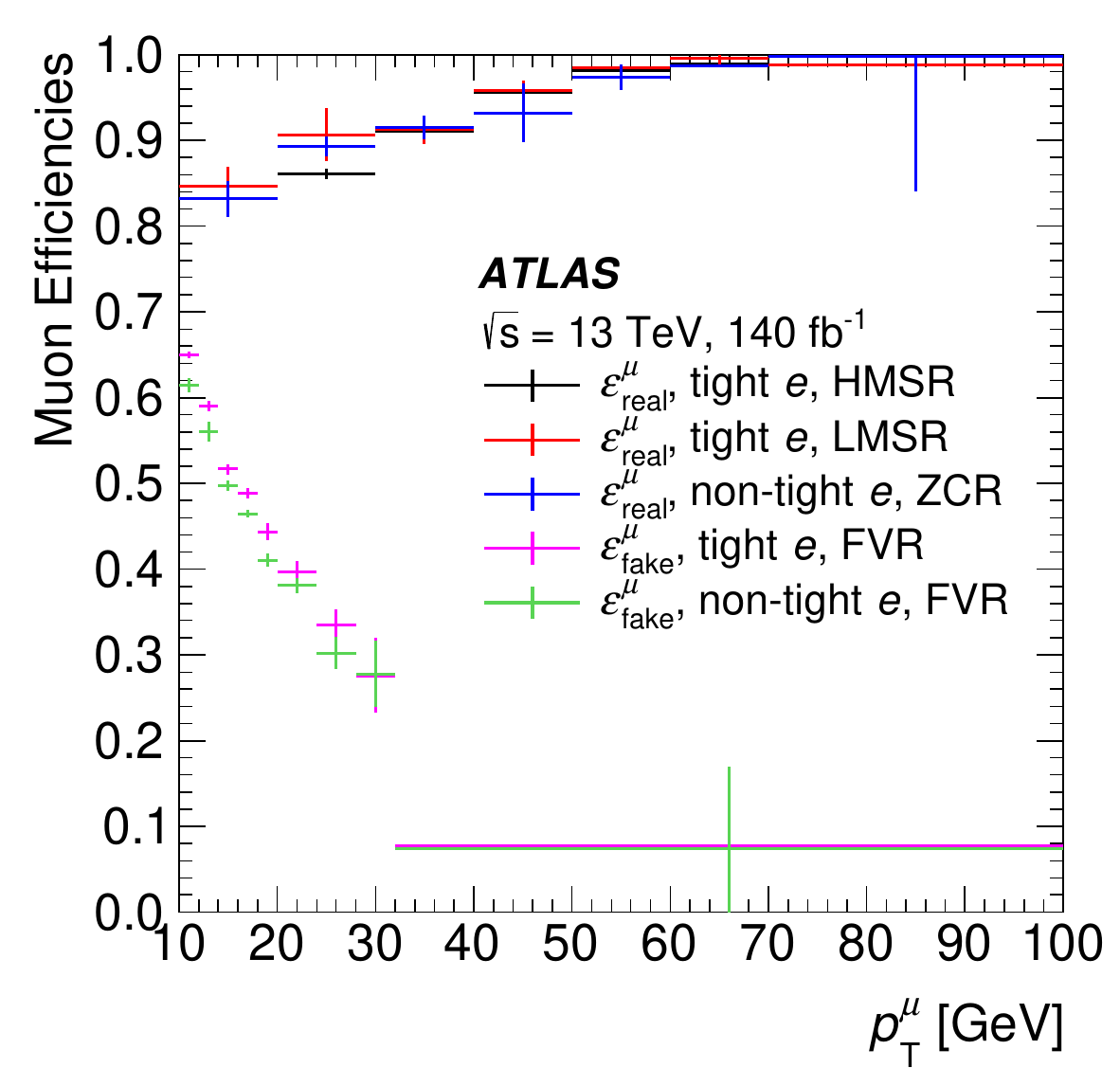}
}
\caption{Fake- and real-lepton efficiencies ($\varepsilon_\mathrm{fake}$ and $\varepsilon_\mathrm{real}$) for (a) electrons and (b) muons with a tight or non-tight accompanying (a) muon and (b) electron, binned in $p_\mathrm{T}^e$ and $p_\mathrm{T}^\mu$, respectively. Real-lepton efficiencies with a tight accompanying lepton are calculated in the high-mass and low-mass SRs (HMSR and LMSR), real-lepton efficiencies with a non-tight accompanying lepton are calculated in the ZCR, and fake-lepton efficiencies are calculated in the FVR.}
\label{fig:efficiencies}
\end{figure}

The real-lepton (fake-lepton) efficiency is defined as the ratio of the number of real (fake) leptons satisfying the tight selection to the number of real (fake) leptons satisfying the baseline selection.
The real-lepton efficiencies are estimated using the \enquote{truth particle} classification from the MC event record, while the fake-lepton efficiencies are estimated from data after subtracting the MC-estimated prompt-lepton background.
Assuming that the electron efficiencies are independent of the muon produced in the same event and vice versa, the conventional matrix method is valid.
In this analysis, however, it was observed that the fake-lepton efficiencies are not entirely independent of the other (fake) lepton in the event.
To account for this, a parameterisation of the real- and fake-lepton efficiencies, based on the tightness of the other lepton, is introduced such that the electron efficiencies depend on the muon's tightness and vice versa:
\begin{align*}
\varepsilon^{e}_\mathrm{real}(\mu) &= \begin{cases}
\cfrac{N^\mathrm{{T}{T}}_\mathrm{RX}}{N^\mathrm{{L}{T}}_\mathrm{RX}} \text{,}\quad \mu{\text{ tight}}\\[10pt]
\cfrac{N^\mathrm{{T}{\overline{T}}}_\mathrm{RX}}{N^\mathrm{{L}{\overline{T}}}_\mathrm{RX}} \text{,}\quad \mu{\text{ non-tight}}
\end{cases},
&\qquad \varepsilon^{\mu}_\mathrm{real}(e) &= \begin{cases}
\cfrac{N^\mathrm{{T}{T}}_\mathrm{XR}}{N^\mathrm{{T}{L}}_\mathrm{XR}} \text{,}\quad e {\text{ tight}}\\[10pt]
\cfrac{N^\mathrm{{\overline{T}}{T}}_\mathrm{XR}}{N^\mathrm{{\overline{T}}{L}}_\mathrm{XR}}  \text{,}\quad e {\text{ non-tight}}
\end{cases},
\end{align*}
\begin{align*}
\varepsilon^{e}_\mathrm{fake}(\mu) &= \begin{cases}
\cfrac{N^\mathrm{{T}{T}}_\mathrm{Data} - N^\mathrm{{T}{T}}_\mathrm{RX}}{N^\mathrm{{L}{T}}_\mathrm{Data} - N^\mathrm{{L}{T}}_\mathrm{RX}}  \text{,}\quad \mu{\text{ tight}}\\[10pt]
\cfrac{N^\mathrm{{T}{\overline{T}}}_\mathrm{Data} - N^\mathrm{{T}{\overline{T}}}_\mathrm{RX}}{N^\mathrm{{L}{\overline{T}}}_\mathrm{Data} - N^\mathrm{{L}{\overline{T}}}_\mathrm{RX}} \text{,}\quad \mu{\text{ non-tight}}
\end{cases},
&\qquad \varepsilon^{\mu}_\mathrm{fake}(e) &= \begin{cases}
\cfrac{N^\mathrm{{T}{T}}_\mathrm{Data} - N^\mathrm{{T}{T}}_\mathrm{XR}}{N^\mathrm{{T}{L}}_\mathrm{Data} - N^\mathrm{{T}{L}}_\mathrm{XR}}  \text{,}\quad e {\text{ tight}}\\[10pt]
\cfrac{N^\mathrm{{\overline{T}}{T}}_\mathrm{Data} - N^\mathrm{{\overline{T}}{T}}_\mathrm{XR}}{N^\mathrm{{\overline{T}}{L}}_\mathrm{Data} - N^\mathrm{{\overline{T}}{L}}_\mathrm{XR}}  \text{,}\quad e {\text{ non-tight}}
\end{cases}.
\end{align*}
Additionally, the efficiencies are binned in lepton \pt as shown in \cref{fig:efficiencies}, revealing that the fake-lepton efficiencies depend on the tightness of the accompanying lepton for low \pt, while the real-lepton efficiencies are statistically compatible.

\begin{figure}[!t] \centering
\subfloat[]{\includegraphics[width=0.49\textwidth]{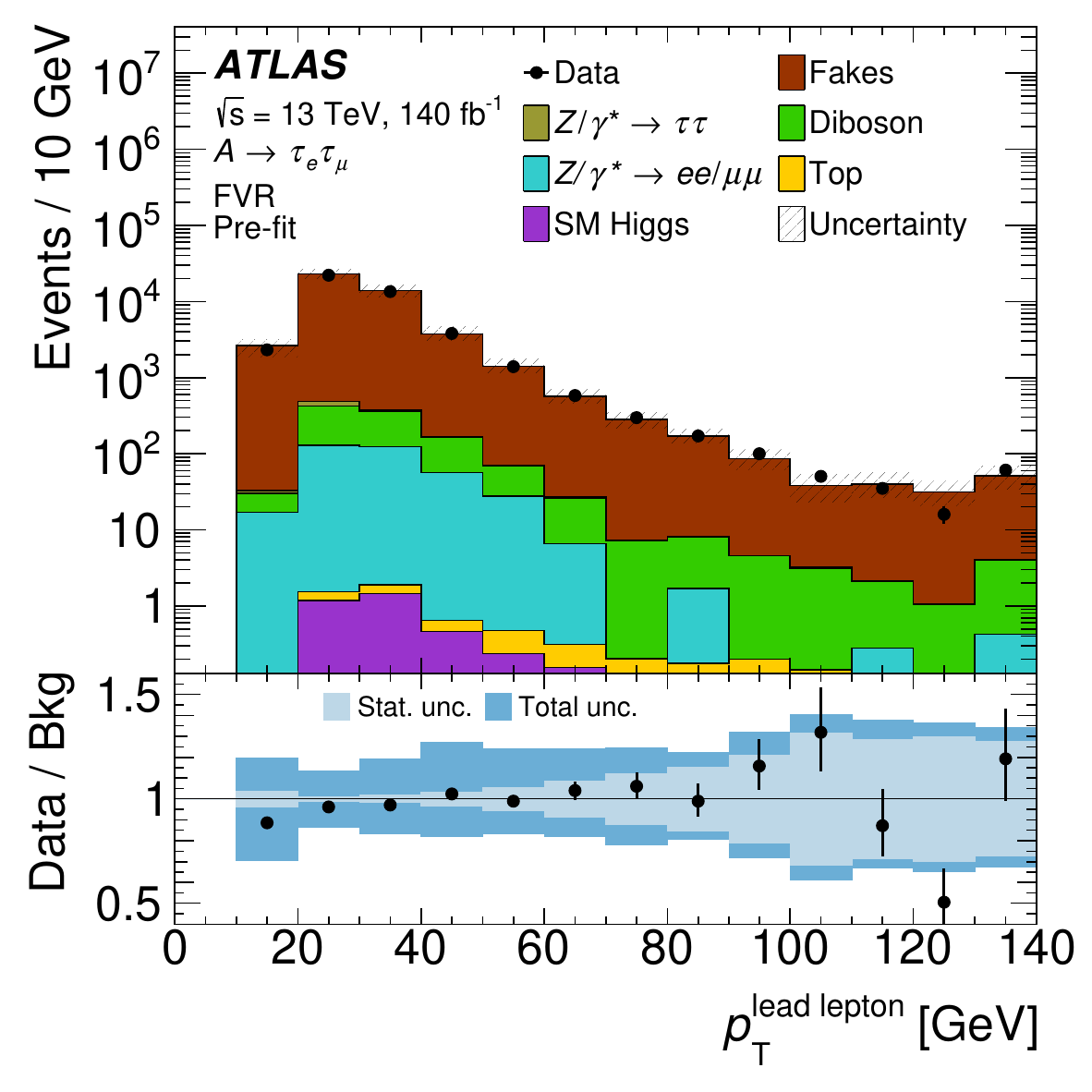}
}
\subfloat[]{\includegraphics[width=0.49\textwidth]{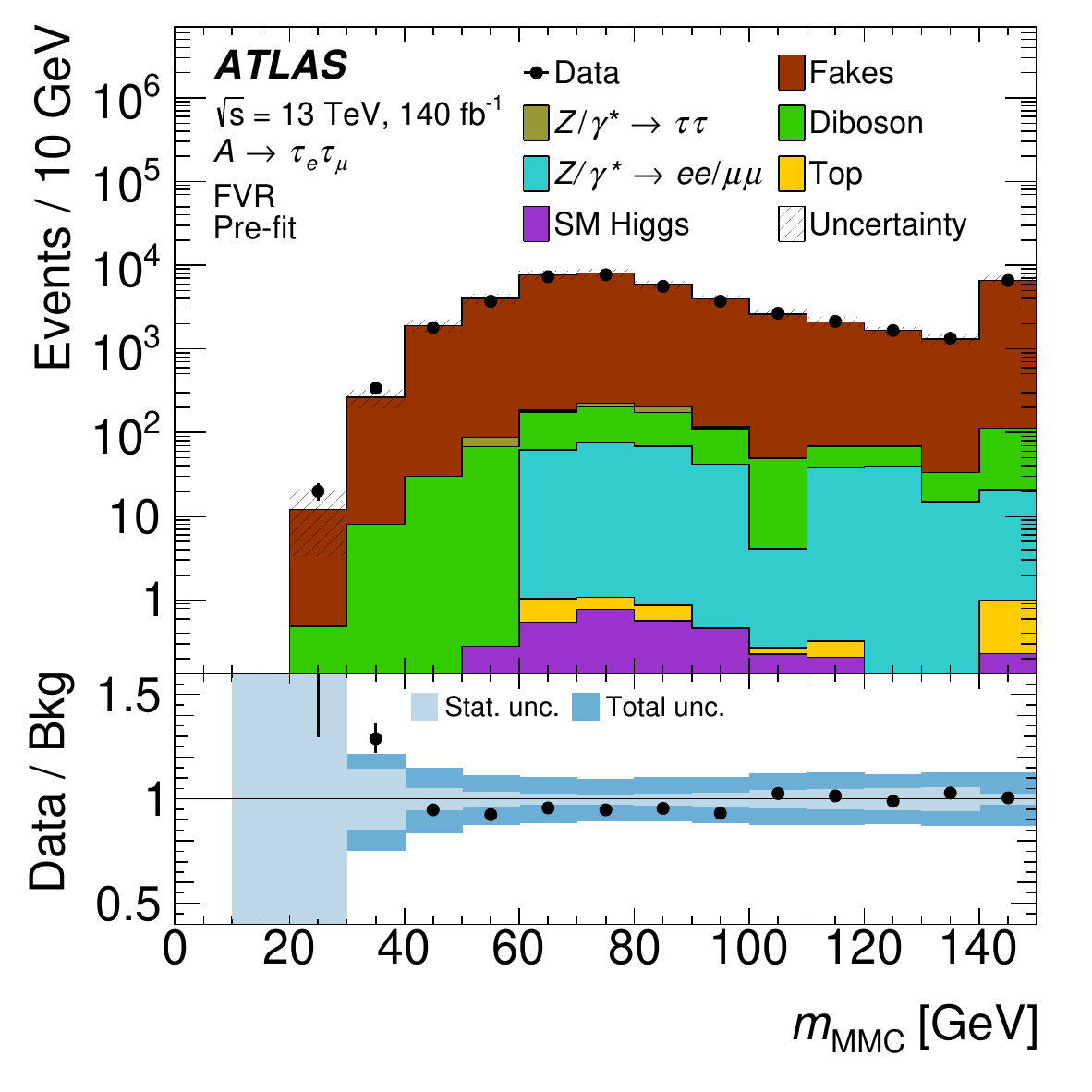}
}
\caption{Pre-fit variable distributions in the FVR after applying all
selection criteria: (a) leading-lepton \pt and (b) \mmc mass estimator.
\commonselectioncaptionFigureVR{}}
\label{fig:sszvr-selection-1}
\end{figure}

When considering real-lepton efficiencies in events with a tight accompanying lepton, it is possible to extract them directly from the SR, as it contains a sufficient number of well-modelled real leptons from prompt-lepton MC background events. Also, different real-lepton efficiencies are calculated for the high-mass SR and low-mass SR, with the $\met$ criterion removed in the latter to increase the event yield.
The real efficiencies for events with a non-tight accompanying lepton are calculated in the ZCR because it has more events than the SR.
To calculate the fake-lepton efficiencies, one needs a region close to the SR but enriched in fake leptons, conditions best satisfied with enough events by the FVR. \cref{fig:sszvr-selection-1} shows the validation of the fake-lepton estimate in the FVR, where the fake-lepton efficiencies are measured. In both the $p_\text{T}^\text{lead lepton}$ and the \mmc distributions, the observed data agrees well with the estimated background within their respective uncertainties.

\subsection{Event yields}
\cref{tab:regions_signal_yields} and \cref{tab:regions_bkg_yields} summarise the event yields for the signal and background predictions, respectively, in the different regions in comparison with those observed in data. Due to the \ztt reweighting the data-to-background ratio in the ZCR equals 1 by definition. In the other control regions, the expected yield is in agreement with the observed number of events within the total uncertainty. The systematic uncertainties are described and estimated in the following section. The purity of the \ztt, top-quark and fake-lepton backgrounds reaches \SI{79}{\%}, \SI{94}{\%} and \SI{97}{\%} in the ZCR, TCR and FVR, respectively. No signal is expected in the FVR and it is negligible in both the ZCR and TCR.
The purity of the signal in the SR depends on the mass hypothesis, and reaches \SI{60}{\%} for masses lower than \SI{60}{\giga\electronvolt} when scaled according to the cross-sections in~\cref{tab:signal_xsec_unc_eff} multiplied by 0.25, the highest currently allowed value of $|\zeta_\mathrm{u}|^2$~\cite{PhysRevD.98.035001}.

\begin{table}[!t]
\begin{center}
\caption[Signal yields]{Pre-fit signal yields in the different analysis regions. The numbers correspond to an integrated luminosity of $\lumi$. The quoted uncertainties are the statistical uncertainties which arise from the finite number of generated events in the simulated samples. No signal is expected in the FVR. The signal cross-sections are calculated to N$^3$LO, as summarised in \cref{tab:signal_xsec_unc_eff}. These cross-section values are multiplied by $0.25=|\zeta_\mathrm{u}|^2=\cot^2\beta$, which is the highest not-already-excluded value of $|\zeta_\mathrm{u}|^2$~\cite{PhysRevD.98.035001}. \newline}
\label{tab:regions_signal_yields}
\small
\providecommand{\xmark}{{\sffamily \bfseries X}}
\providecommand\rotatecell[2]{\rotatebox[origin=c]{#1}{#2}}
\renewcommand{\arraystretch}{1.5}
\begin{tabular}{ c |*{2}{r@{$\,\pm\,$}l} |*{2}{r@{$\,\pm\,$}l}}
\hline\hline
$m_A$ [\si{\giga\electronvolt}] & \multicolumn{2}{c}{{ZCR}} & \multicolumn{2}{c|}{{TCR}} &  \multicolumn{2}{c}{{Low-mass SR}} & \multicolumn{2}{c}{{High-mass SR}} \\
\hline\hline
20 & $\numRF[1]{0.0}   $ & $\numRF[1]{0.0}   $ & $\numRF[3]{15.4}$ & $\numRF[2]{3.0}  $ & $\numRF[4]{1354.6}$ & $\numRF[2]{27.2}  $ & \multicolumn{2}{c}{-} \\
25 & $\numRF[1]{0.3}   $ & $\numRF[1]{0.3}   $ & $\numRF[3]{30.8}$ & $\numRF[2]{3.3}  $ & $\numRF[4]{2326.0}$ & $\numRF[2]{28.0}  $ & \multicolumn{2}{c}{-} \\
30 & $\numRF[2]{4.7}   $ & $\numRF[2]{1.8}   $ & $\numRF[3]{39.6}$ & $\numRF[2]{5.3}  $ & $\numRF[4]{2865.1}$ & $\numRF[2]{44.4}  $ & \multicolumn{2}{c}{-} \\
40 & $\numRF[3]{313.8} $ & $\numRF[2]{16.9}  $ & $\numRF[3]{59.4}$ & $\numRF[2]{7.6}  $ & $\numRF[4]{2940.1}$ & $\numRF[2]{52.4}  $ & \multicolumn{2}{c}{-} \\
50 & $\numRF[4]{1705.5}$ & $\numRF[2]{42.8}  $ & $\numRF[3]{51.2}$ & $\numRF[2]{7.5}  $ & $\numRF[4]{2292.9}$ & $\numRF[2]{50.7}  $ & \multicolumn{2}{c}{-} \\
60 & $\numRF[4]{4396.2}$ & $\numRF[2]{71.1}  $ & $\numRF[3]{39.6}$ & $\numRF[2]{7.1}  $ & $\numRF[4]{1515.6}$ & $\numRF[2]{42.6}  $ & \multicolumn{2}{c}{-} \\
70 & $\numRF[4]{7242.0}$ & $\numRF[2]{90.6}  $ & $\numRF[3]{53.6}$ & $\numRF[2]{8.1}  $ & $\numRF[4]{1002.4}$ & $\numRF[2]{34.6}  $ & \multicolumn{2}{c}{-} \\
75 & $\numRF[4]{8083.0}$ & $\numRF[2]{74.5}  $ & $\numRF[3]{53.9}$ & $\numRF[2]{6.5}  $ & $\numRF[3]{773.5} $ & $\numRF[2]{23.4}  $ & $\numRF[4]{2311.9}$ & $\numRF[2]{40.5}$ \\

80  & $\numRF[4]{8519.7}$ & $\numRF[2]{95.5}  $ & $\numRF[3]{49.9}$ & $\numRF[2]{7.5}  $ & \multicolumn{2}{c}{-}    & $\numRF[4]{2010.5}$ & $\numRF[2]{47.1}$ \\
85  & $\numRF[4]{8706.2}$ & $\numRF[2]{76.0}  $ & $\numRF[3]{39.1}$ & $\numRF[2]{5.4}  $ & \multicolumn{2}{c}{-}    & $\numRF[4]{1771.4}$ & $\numRF[2]{34.8}$ \\
90  & $\numRF[4]{8710.3}$ & $\numRF[2]{92.7}  $ & $\numRF[3]{39.6}$ & $\numRF[2]{6.3}  $ & \multicolumn{2}{c}{-}    & $\numRF[4]{1512.0}$ & $\numRF[2]{39.2}$ \\\hline\hline
\end{tabular}
\renewcommand{\arraystretch}{1.0}
\end{center}
\end{table}

\begin{table}[!t]
\begin{center}
\caption[Background yields]{Pre-fit background event yields in the different analysis regions. The numbers correspond to an integrated luminosity of $\lumi$. The quoted uncertainties are the statistical uncertainties which arise from the finite number of generated events in the simulated samples and from the statistical uncertainties of the measured data as input for the fake-lepton background. The values in parentheses represent the fraction of the corresponding background relative to the total background in each region. For the $Z/\gamma^{*}\to \ell^+\ell^-$ background, $\ell^+\ell^-$ includes $e^+e^-$ and $\mu^+\mu^-$.
}
\label{tab:regions_bkg_yields}
\small
\providecommand{\xmark}{{\sffamily \bfseries X}}
\providecommand\rotatecell[2]{\rotatebox[origin=c]{#1}{#2}}
\renewcommand{\arraystretch}{1.5}
\begin{tabular}{ l | r@{$\,\pm\,$}l  r@{$\,\pm\,$}l  r@{$\,\pm\,$}l  | r@{$\,\pm\,$}l  r@{$\,\pm\,$}l } %
\hline\hline
& \multicolumn{2}{c}{ZCR} & \multicolumn{2}{c}{FVR} & \multicolumn{2}{c |}{TCR} & \multicolumn{2}{c}{Low-mass SR} & \multicolumn{2}{c}{High-mass SR}  \tabularnewline
\hline\hline
Fakes  & $\numRF[4]{59677.3}$ & $ \numRF[2]{691.1}$ & $\numRF[4]{44850.7}$ & $ \numRF[2]{434.2}$ & $\numRF[3]{212.8}$ & $ \numRF[2]{49.5}$ & $\numRF[3]{315.7}$ & $ \numRF[2]{44.5}$ & $\numRF[3]{495.3}$ & $ \numRF[2]{55.4} $ \tabularnewline
\rule{0pt}{6pt}  & \multicolumn{2}{c}{($\numRF[2]{18.0}\%)$}  & \multicolumn{2}{c}{($\numRF[2]{97.4}\%)$}  & \multicolumn{2}{c |}{$(\numRF[2]{3.5}\%)$}  &  \multicolumn{2}{c}{($\numRF[2]{16.5}\%)$}  & \multicolumn{2}{c}{($\numRF[2]{9.0}\%)$}  \tabularnewline

$Z/\gamma^{*}\to\tau^+\tau^-$ & $\numRF[4]{262664.1}$ & $ \numRF[2]{1828.4}$ & $\numRF[2]{81.970}$ & $ \numRF[2]{27.4}$ & $\numRF[4]{116.1}$ & $ \numRF[2]{4.0}$ & $\numRF[4]{1209.7}$ & $ \numRF[2]{36.2}$ & $\numRF[4]{3700.6}$ & $ \numRF[2]{46.0} $ \tabularnewline
\rule{0pt}{6pt}  & \multicolumn{2}{c}{($\numRF[2]{79.2}\%)$}  & \multicolumn{2}{c}{($\numRF[2]{0.178}\%)$}  & \multicolumn{2}{c |}{$(\numRF[2]{1.9}\%)$}  & \multicolumn{2}{c}{($\numRF[2]{63.4}\%)$} & \multicolumn{2}{c}{($\numRF[2]{67.4}\%)$} \tabularnewline

Diboson & $\numRF[4]{4552.0}$ & $ \numRF[2]{25.5}$ & $\numRF[4]{747.3}$ & $ \numRF[2]{7.4}$ & $\numRF[4]{10.823}$ & $ \numRF[2]{0.503} $ & $\numRF[4]{139.1}$ & $ \numRF[2]{2.4} $ & $\numRF[4]{449.2}$ & $ \numRF[2]{4.4} $ \tabularnewline
\rule{0pt}{6pt}  & \multicolumn{2}{c}{($\numRF[2]{1.4}\%)$}  & \multicolumn{2}{c}{($\numRF[2]{1.6}\%)$}  & \multicolumn{2}{c |}{$(\numRF[2]{0.180}\%)$}  & \multicolumn{2}{c}{($\numRF[2]{7.3}\%)$}  & \multicolumn{2}{c}{($\numRF[2]{8.2}\%)$} \tabularnewline

$Z/\gamma^{*}\to \ell^+\ell^-$ & $\numRF[3]{467.6}$ & $ \numRF[2]{71.4} $ & $\numRF[3]{353.7}$ & $ \numRF[2]{67.8} $ & $\numRF[2]{0.283}$ & $ \numRF[2]{0.116} $ & $\numRF[2]{4.0}$ & $ \numRF[2]{1.7} $ & $\numRF[2]{9.8}$ & $ \numRF[2]{3.3} $ \tabularnewline
\rule{0pt}{6pt}  & \multicolumn{2}{c}{($\numRF[2]{0.141}\%)$}  & \multicolumn{2}{c}{($\numRF[2]{0.768}\%)$}  & \multicolumn{2}{c |}{($<0.1\%$)}  & \multicolumn{2}{c}{($\numRF[2]{0.208}\%)$}  & \multicolumn{2}{c}{$(\numRF[2]{0.179}\%)$} \tabularnewline

Top & $\numRF[4]{3327.2}$ & $ \numRF[2]{21.7}$ & $\numRF[3]{2.024}$ & $ \numRF[2]{0.509}$ & $\numRF[4]{5652.9}$ & $ \numRF[2]{27.7} $ & $\numRF[4]{162.9}$ & $ \numRF[2]{4.7} $ & $\numRF[4]{611.4}$ & $ \numRF[2]{9.3} $ \tabularnewline
\rule{0pt}{6pt}  & \multicolumn{2}{c}{($\numRF[2]{1.0}\%)$} & \multicolumn{2}{c}{($<0.1\%$)} & \multicolumn{2}{c |}{$(\numRF[2]{94.2}\%)$}  & \multicolumn{2}{c}{($\numRF[2]{8.5}\%)$}  & \multicolumn{2}{c}{$(\numRF[2]{11.1}\%)$} \tabularnewline

SM Higgs & $\numRF[5]{1108.7}$ & $ \numRF[2]{2.6} $ & $\numRF[3]{3.634}$ & $ \numRF[2]{0.176} $ & $\numRF[3]{5.794}$ & $ \numRF[2]{0.226} $ & $\numRF[4]{76.861}$ & $ \numRF[2]{0.789} $ & $\numRF[4]{227.4}$ & $ \numRF[2]{1.3} $ \tabularnewline
\rule{0pt}{6pt}  & \multicolumn{2}{c}{($\numRF[2]{0.334}\%)$} & \multicolumn{2}{c}{($<0.1\%$)} & \multicolumn{2}{c |}{$(\numRF[2]{0.097}\%)$} & \multicolumn{2}{c}{($\numRF[2]{4.0}\%$)}  & \multicolumn{2}{c}{$(\numRF[2]{4.1}\%)$} \tabularnewline

\hline
Total background & $\numRF[4]{331796.9}$ & $ \numRF[2]{1956.3}$ & $\numRF[4]{46039.3}$ & $ \numRF[2]{440.3}$ & $ \numRF[4]{5998.7}$ & $ \numRF[2]{56.8}$ & $\numRF[4]{1908.2}$ & $ \numRF[2]{57.6}$ & $\numRF[4]{5493.7}$ & $ \numRF[2]{72.8} $ \tabularnewline
\hline
Data &  \multicolumn{2}{l}{$\numRF[6]{331797}$} & \multicolumn{2}{l}{$\numRF[5]{44587}$} & \multicolumn{2}{l |}{\phantom{.}$\numRF[4]{6227}$} & \multicolumn{2}{l}{\phantom{.}$\numRF[4]{1987}$} & \multicolumn{2}{l}{\phantom{.}$\numRF[4]{6119}$} \tabularnewline
\hline\hline
\end{tabular}
\renewcommand{\arraystretch}{1.0}
\end{center}
\end{table}


%

\section{Systematic uncertainties}
\label{sec:systematics}
%

Systematic uncertainties are categorised into those arising from the measurements (experimental uncertainties) and those due to the modelling of various processes (theoretical uncertainties), where the latter are specific to a single process or a group of processes.
In contrast, the experimental uncertainties depend on simulations of the detector and reconstruction of objects, and are therefore independent of the process.
A nuisance parameter (NP) with a Gaussian prior is assigned to each systematic uncertainty.
Additionally, smoothing and pruning procedures are used to reduce the impact of statistical fluctuations and remove the smallest systematic uncertainties evaluated by performing various \enquote{systematic variations}.
The shape and normalisation components of the systematic uncertainties are dropped if they do not exceed \SI{0.5}{\%} for a given process (in the case of the shape uncertainties, if no bin shows an effect greater than \SI{0.5}{\%}).

\subsection{Experimental uncertainties}
The uncertainty in the combined 2015--2018 integrated luminosity\footnote{The luminosity uncertainty is only included in the limit calculation and not in the variables' distribution plots.} is 0.83\%~\cite{DAPR-2021-01}, obtained using the LUCID-2 detector~\cite{LUCID2} for the primary luminosity measurements.
A systematic uncertainty is assigned to the reweighting procedure, based on the mean number of $pp$ interactions per bunch crossing, used to correct the pile-up profile in MC simulation to match the data.
For leptons, systematic uncertainties are assigned to the scale factors applied to correct for all differences between data and simulation that come from the electron and muon identification, reconstruction, isolation and trigger performance~\cite{TRIG-2018-05,TRIG-2018-01,EGAM-2018-01,MUON-2018-03} or from the lepton momentum scale and resolution.
For jets, systematic uncertainties from the mismodelling of jet energy scale and resolution effects~\cite{JETM-2018-05}, jet vertex-tagging efficiencies~\cite{PERF-2014-03}, and jet flavour-tagging efficiencies~\cite{PERF-2016-05,FTAG-2020-08,FTAG-2019-02} are considered. For the missing transverse momentum, uncertainties from the scale and resolution of the track-based \enquote{soft term} are considered~\cite{JETM-2020-03}. The latter are one-sided variations. For the \met \enquote{hard terms}, uncertainties are already covered by electron, muon and jet uncertainties.

Other kinds of uncertainties are considered for the fake- and non-prompt-lepton backgrounds.
First, the statistical uncertainties of the real- and fake-lepton efficiencies for electrons and muons are
propagated as a single systematic uncertainty for each \pt bin.
Also, the dependence of the efficiencies on variables other than the lepton \pt need to be accounted for.
However, due to the limited number of data events, it is not possible to split the efficiencies into more components,
so systematic uncertainties for the unaccounted dependencies of the efficiencies are included by parameterising them in $\eta$, $n_\mathrm{jets}$ and $\met$.
For the $\eta$ dependence, three different bins of absolute $\eta$ values are introduced: $[0,0.8)$, $[0.8,1.7)$ and $[1.7,2.7]$.
The dependence of the efficiencies on the number of jets, $n_\mathrm{jets}$, is also modelled, using four $n_\mathrm{jets}$ values $\{0,1,2, {\ge}3\}$ for the fake-lepton efficiencies and $\{1,2,3, {\ge}4\}$ for the real-lepton efficiencies, as there are no events without jets in the signal regions.
For the $\met$ dependence, four different bins of $\met$ values are used:\ $[0,10)\,\si{\giga\electronvolt}$, $[10,20)\,\si{\giga\electronvolt}$, $[20,30)\,\si{\giga\electronvolt}$ and $[30,\infty)\,\si{\giga\electronvolt}$.
For all those parameterisation uncertainties, the weighted standard deviations of the efficiencies are calculated and used as systematic uncertainties.
Furthermore, the region where the efficiencies are measured and the SR could have different compositions.
To account for this difference, a composition uncertainty is included by calculating the efficiencies in regions with the same requirements as the signal regions, but with the leptons having the same charge.
Also, the \mmc and $\met$ criteria are omitted to increase the number of events, as the $\met$ parameterisation is already accounted for by a dedicated uncertainty.
Both the parameterisation and composition uncertainties are decorrelated across the regions for the fit because the regions use independent data, and allowing correlations would overly constrain the corresponding NPs.
In some cases, a systematic uncertainty may arise from the uncertainty in the MC-event subtraction when calculating the fake-lepton efficiencies.
For this analysis, the MC-event contamination in the FVR, where the fake-lepton efficiencies are measured, is only \SI{3}{\%}.
Thus, the uncertainties coming from the MC-event contamination are not considered in this analysis.

The most significant experimental uncertainties stem from the fake-lepton background estimation, flavour tagging, and muon efficiencies and calibration.

\subsection{Theoretical uncertainties}

Systematic uncertainties of the signal and background production cross-sections arise from various sources: missing higher-order (MHO) corrections, the PDFs, and the chosen input parameter values. The relative uncertainties in the inclusive cross-sections for backgrounds are \SI{1.9}{\%} (\SI{5}{\%})~\cite{STDM-2015-03} for \ztt with $m_{\ell\ell}$ above (below) $\SI{66}{\giga\electronvolt}$, \SI{4.4}{\%}~\cite{Czakon:2011xx, Botje:2011sn,Martin:2009bu,Gao:2013xoa,Ball:2012cx} for top-quark processes (dominated by \ttbar), and an envelope of \SI{7.1}{\%}~\cite{STDM-2017-24,STDM-2017-03,STDM-2018-03} for diboson processes. The relative uncertainties in the signal cross-sections are only taken into account for the model-dependent interpretation, with the estimated values ranging from \SI{6.5}{\%} to \SI{11.2}{\%}.

Generator systematic uncertainties for \ztt and diboson processes arise mainly from the renormalisation and factorisation scales ($\mu_\mathrm{R}$ and $\mu_\mathrm{F}$), as well as \alphas uncertainties, PDF parameter value choices, and uncertainties in the resummation (QSF) and CKKW matching scales. The related event samples were produced using \SHERPA[2.2.1] and 2.2.2, respectively. Systematic uncertainties associated with renormalisation and factorisation scale, PDF and \alphas variations are assessed using corresponding event weights. All uncertainties are calculated in the high-mass SR and ZCR. The latter are applied in the ZCR and FVR. In all other regions, the ones calculated in the high-mass SR are used.
For the QCD scale uncertainties, a 7-point variation is performed by varying $\mu_\mathrm{R}$ and $\mu_\mathrm{F}$ by factors of 2 and 0.5 in different combinations. For each bin, the uncertainty is given by the maximum positive and negative changes due to the variations.
The scale uncertainty for the \ztt process is decorrelated between the regions because of its largely different structure in the high-mass SR and the ZCR and to account for potential problems in the extrapolation of $n_\mathrm{jets}$ weights from the ZCR described in \cref{sec:promptbackground}.
Using the baseline PDF, \NNPDF[3.0nlo] with a strong coupling value of $\alphas(m_Z ) = 0.118$, as the nominal one, the standard deviation of 100 PDF variations is determined and taken as a symmetric uncertainty. Also, \alphas is varied from 0.118 by $\pm 0.001$. The \CT[14nnlo]~\cite{Dulat:2015mca} and \MMHT[nnlo] PDF sets with associated \SI{68}{\%} confidence-level uncertainty bands~\cite{Harland-Lang:2014zoa} are compared with the baseline PDF to cross-check the results.
Uncertainties corresponding to CKKW and QSF scale variations are evaluated only for \ztt, via a parameterisation method. For the CKKW matching, the nominal scale used in calculations of the overlap between jets from the ME and parton shower is \SI{20}{\giga\electronvolt} and is varied between \SI{15}{\giga\electronvolt} and \SI{30}{\giga\electronvolt}. The soft-gluon emission resummation scale is varied by factors of 0.5 and 2 from its nominal value.
All of the \ztt and diboson generator uncertainties are calculated in bins of \mmmc.

Additional systematic uncertainties are included for the \ztt reweighting, described in \cref{sec:promptbackground}, by propagating the statistical uncertainties of MC events and data events, as well as experimental and generator systematic uncertainties of MC events, to the weights.
The resulting uncertainties cover a reweighting value of unity in most $n_\mathrm{jets}$ bins.
The NPs are decorrelated across the regions to avoid double-counting the propagation of effects from the ZCR (where the weights were determined) to the SR.

The uncertainty in the modelling of top-quark processes has contributions from different sources. The uncertainties from each source are estimated by comparing the nominal sample (\POWHEGBOX[v2] at NLO with \NNPDF[3.0nlo] and $\hdamp=1.5\,m_\mathrm{top}$; \PYTHIA[8.230] with the A14 tune and \NNPDF[2.3lo]; and \EVTGEN[1.6.0]) with multiple variations. The uncertainties are evaluated in a modified TCR. Compared to the nominal TCR, the requirement on the number of $b$-tagged jets is loosened to ${\geq}1$~$b$-tagged jet in order to accept more events.\footnote{Systematic variations are performed using simulated top-quark processes only and are thus not affected by a possible signal contamination in an extended TCR.}
The impact of using a different parton shower and hadronisation model is evaluated by replacing \PYTHIA with \HERWIG[7.13]~\cite{Bahr:2008pv,Bellm:2015jjp} and the \MMHT[lo] PDF set; for the impact of using different ME generation and matching to the parton shower, \MGNLO[2.6.0]~\cite{Alwall:2014hca} is used in place of \POWHEGBOX; and to estimate the impact of changing the \hdamp parameter value, a sample with $\hdamp=3\,m_\mathrm{top}$ is used. Separate alternative datasets were simulated for each of the above variations.
Initial-state radiation (ISR) uncertainties are estimated by halving and doubling $\mu_\mathrm{R}$ and $\mu_\mathrm{F}$, and using the A14 tune's Var3cUp and Var3cDown variations corresponding to the variation of \alphas for ISR. For final-state radiation (FSR), halving and doubling the renormalisation scale for parton shower emissions gives an estimate of the relevant uncertainty. These \alphas and scale variations use alternative event weights recorded in the nominal simulation datasets.
All of the top-quark generator uncertainties are calculated in bins of \mmmc.

Uncertainties in signal modelling arise from the following sources: MHO corrections, the PDFs, the colour reconnection parameters and \alphas. Furthermore, uncertainties from the modelling of ISR and FSR, multi-parton interactions (MPI), parton showers and hadronisation are considered.
They are estimated from ratios of nominal-sample event acceptance to variation-sample event acceptance, binned in $p_\mathrm{T}^A$.
Uncertainties arising from the modelling of ISR, FSR, MPI and colour reconnection are estimated by adding in quadrature the effects of the Var1, Var2, Var3a, Var3b and Var3c up and down eigenvariations of \Pythia[8]'s A14 tune.
The scale uncertainties due to MHO corrections are determined from event weights given for 7-point combinations of different $\mu_\mathrm{R}$ and $\mu_\mathrm{F}$ values~\cite{deFlorian:2016spz}.
Analogously, the PDF and \alphas uncertainties are estimated from variations given in the form of event weights, with the \NNPDF[2.3lo] PDF as the baseline, for two \alphas values $0.119$ and $0.130$, and for 100 variations of the PDF set.
For every $m_A$ value and each systematic variation, about 5~million events were generated to reduce statistical effects. Finally, a smoothing algorithm is applied, taking into account neighbouring mass hypotheses, to reduce remaining statistical fluctuations.


%

\section{Statistical model and results}
%
%

The signal strength $\mu$, defined as a multiplicative factor to the cross-section of the hypothetical signal, is obtained from the statistical analysis of the data using a binned likelihood function based on a product of Poisson probability terms.
Signal and background predictions depend on systematic uncertainties, which are parameterised as NPs and constrained using Gaussian functions.
Normalisation factors for the \ztt and top-quark backgrounds are also included but are not constrained.
The binned likelihood function is constructed in bins of \mmmc.
The binning was chosen in a way that ensures the statistical uncertainty per bin does not exceed \SI{10}{\%} and that there are at least ten background events per bin, while requiring a minimum bin width of \SI{2}{\giga\electronvolt}.
Additionally, at least one fake-lepton event per bin is required.
This is technically necessary in order to avoid having bins with negative expected background contributions or background expectations close to zero with large uncertainties in the statistical analysis.
The fit model consists of three regions: the low-mass or high-mass SR, depending on the mass hypothesis, and the two control regions, TCR and ZCR.
The two control regions allow the extraction of a normalisation factor for the top-quark and \ztt backgrounds and constrain some of the systematic uncertainties.
To account for shape effects in the TCR and ZCR, the regions are split into five and six \mmc bins, respectively.
The FVR is not used in the fit.
The statistical uncertainties of the simulated background and signal, and those of the fake-lepton background histograms, are modelled by dedicated NPs, which use Poisson constraint terms.
They are introduced separately for the signals and the combined backgrounds.

\begin{figure}[!t] \centering
\subfloat[]{\includegraphics[width=0.49\textwidth]{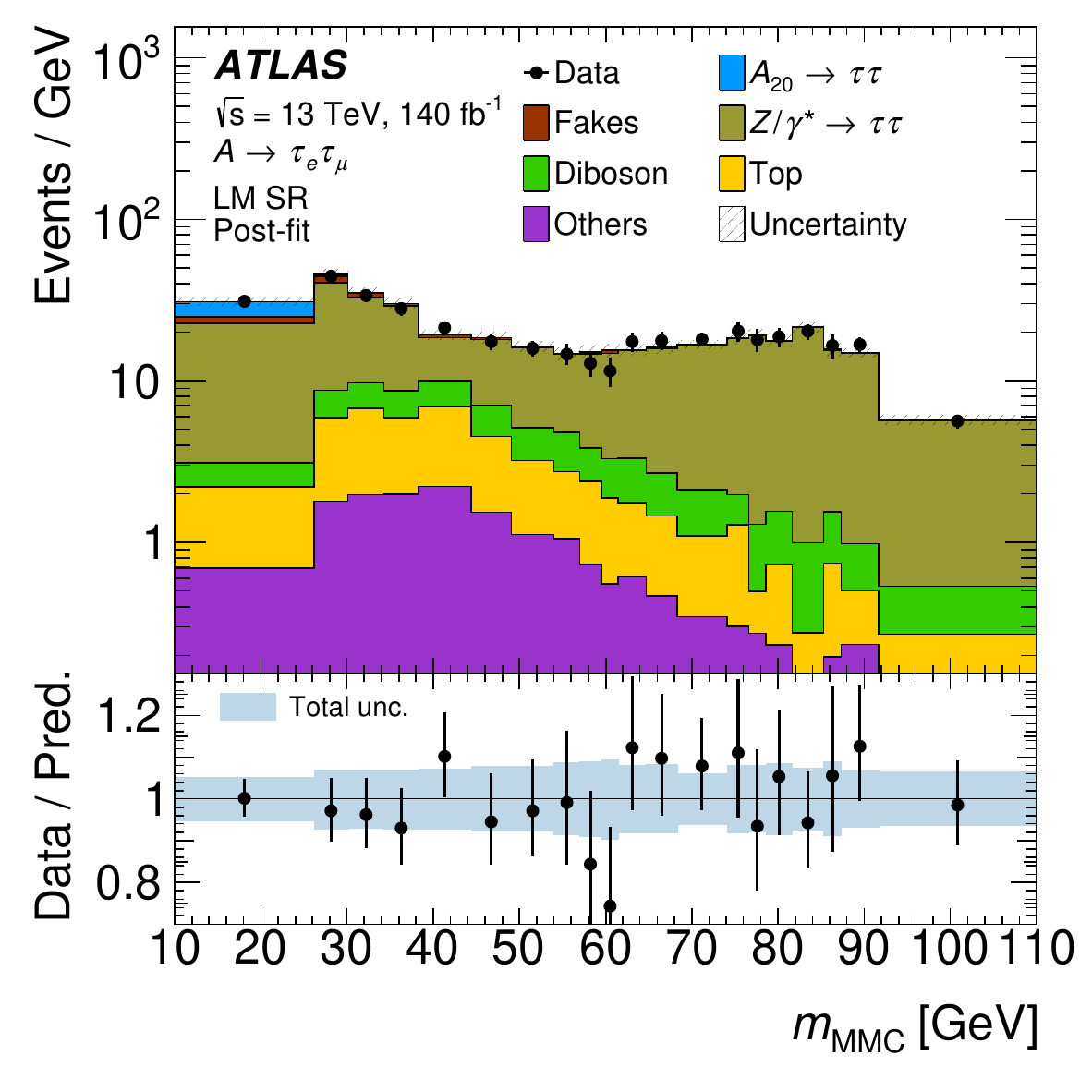}
}
\subfloat[]{\includegraphics[width=0.49\textwidth]{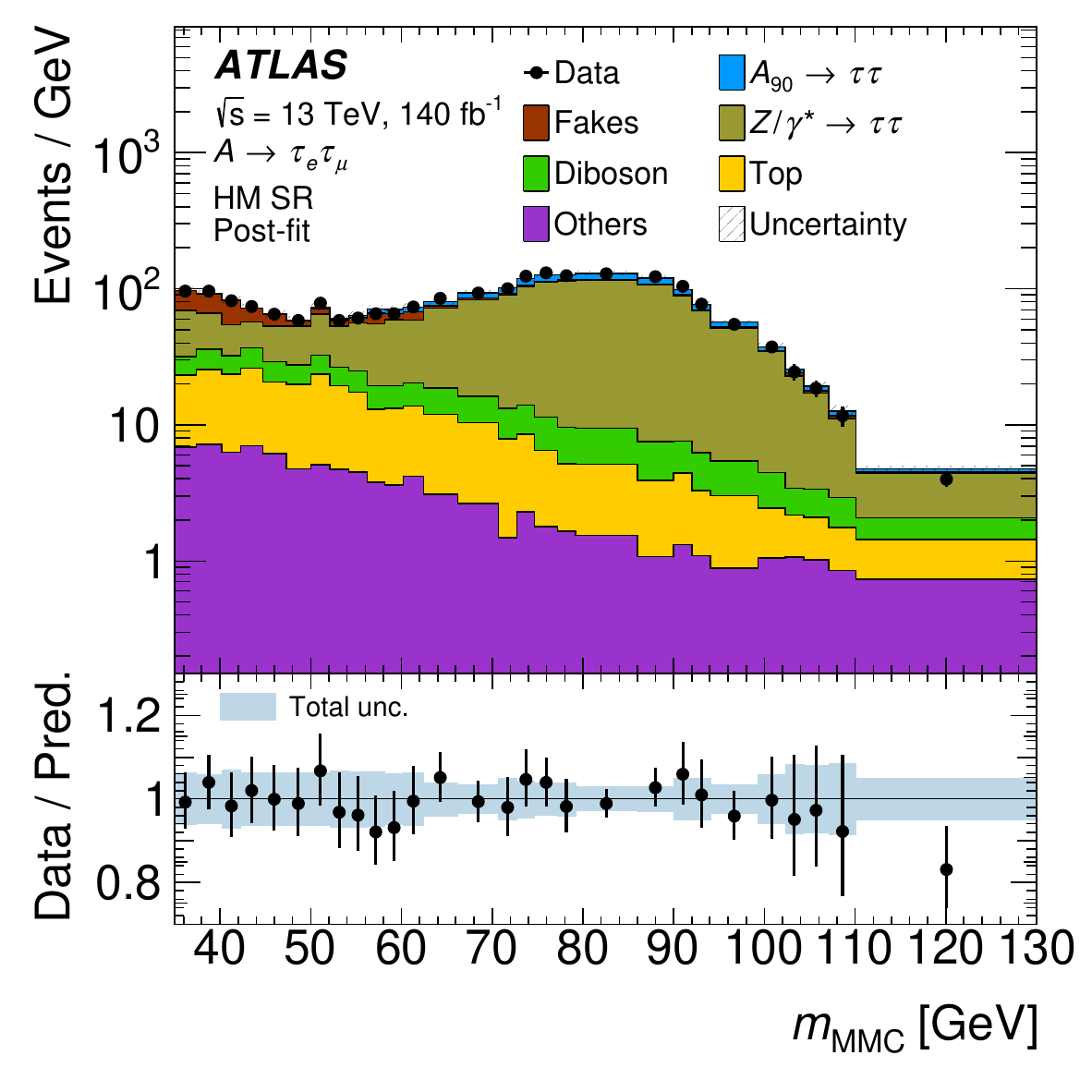}
}
\caption{Post-fit $m_\mathrm{MMC}$ distributions in (a) the low-mass SR for the $m_A=20\,\si{\giga\electronvolt}$ signal mass hypothesis and (b) the high-mass SR for the $m_A=90\,\si{\giga\electronvolt}$ signal mass hypothesis. The fits are performed under the signal-plus-background hypothesis using the signal with the respective $m_A$.
\commonselectioncaptionPostfitOthers{}
\commonselectioncaptionOverflowSR{}
\commonselectioncaptionRatioPostfit{}
}
\label{fig:postFitMMC}
\end{figure}

The fit is performed separately for each signal mass hypothesis. The final \mmc distributions after the fit are shown in~\cref{fig:postFitMMC} for the \SI{20}{\giga\electronvolt} mass hypothesis from the low-mass SR and for the \SI{90}{\giga\electronvolt} mass hypothesis from the high-mass SR, including both the statistical and systematic uncertainties.
Although the signal contributions to the control regions are negligible, they are taken into account properly in the fit.

The largest uncertainties in the SR, measured by their grouped impact on the fit result are listed in \cref{tab:relative_grouped_impact_2090}.
The dominant systematic uncertainties originate from the fake-lepton estimation, the \ztt modelling, the \ztt reweighting and the \ztt MC sample size. Data statistical uncertainties play only a minor direct role; however, they play an important indirect role through the statistical uncertainties of the fake-lepton estimate.

\begin{table}
\centering
\caption{Breakdown of the relative contributions to the uncertainty of the extracted signal strength $\mu$ in the likelihood fit to data. The contributions are obtained by fixing the relevant NPs to their post-fit values in the likelihood fit. The relative impact is determined as the square-root of the difference of the squares of the nominal uncertainty and the obtained uncertainty, divided by the nominal uncertainty. The sum in quadrature of the individual components differs from the total uncertainty because of correlations between uncertainties in the different groups. The uncertainty from data statistical uncertainties is determined from fits with all NPs fixed to their post-fit values. The background statistical uncertainty includes the statistical uncertainty of the fake-lepton background determination and uncertainties due to background MC statistics. Only NP groups with a relative impact larger than \SI{5}{\%} are listed.}
\label{tab:relative_grouped_impact_2090}
\begin{tabular}{lr|lr}
\hline\hline
\multicolumn{2}{c|}{\rule{0pt}{11pt}$m_A=\SI{20}{\giga\electronvolt}$} & \multicolumn{2}{c}{\rule{0pt}{11pt}$m_A=\SI{90}{\giga\electronvolt}$} \\
\rule{0pt}{11pt} Category & \multicolumn{1}{l|}{Relative} & Category & \multicolumn{1}{l}{Relative}  \\
\rule{0pt}{11pt}  & \multicolumn{1}{l|}{contribution} &  &  \multicolumn{1}{l}{contribution} \\		\rule{0pt}{11pt}  & \multicolumn{1}{l|}{to $\Delta\mu$} &  &  \multicolumn{1}{l}{to $\Delta\mu$} \\\hline\hline
\rule{0pt}{11pt}\textbf{Data statistical} & 42\% & \textbf{Data statistical} & 10\% \\
\rule{0pt}{11pt}\textbf{Systematic} & 91\% & \textbf{Systematic} & 99\% \\
\rule{0pt}{11pt} ~~~Background statistical & 81\% & ~~~Background statistical & 67\% \\
\rule{0pt}{11pt} ~~~~~~\ztt statistical & 75\% & ~~~~~~\ztt statistical & 48\% \\
\rule{0pt}{11pt} ~~~~~~Fake-lepton statistical & 30\% & ~~~~~~Fake-lepton statistical & 47\% \\
\rule{0pt}{11pt} ~~~~~~Other background statistical & 7\% & ~~~~~~Other background statistical & 8\% \\
\rule{0pt}{11pt} ~~~Fake-lepton systematic & 37\% & ~~~\ztt reweighting & 55\% \\
\rule{0pt}{11pt} ~~~Signal modelling & 11\% & ~~~Signal statistical & 51\% \\
\rule{0pt}{11pt} ~~~\ztt modelling & 10\% & ~~~Fake-lepton systematic & 47\% \\
\rule{0pt}{11pt} ~~~Muon efficiencies & 8\% & ~~~\ztt modelling & 45\% \\
\rule{0pt}{11pt} ~~~Diboson modelling & 8\% & ~~~\ztt and $t\bar{t}$ normalisation & 34\% \\
\rule{0pt}{11pt} ~~~Flavour tagging & 5\% & ~~~$t\bar{t}$ modelling & 14\% \\
\rule{0pt}{11pt} ~~~Signal statistical & 5\% & ~~~Flavour tagging & 14\% \\
& & ~~~Muon calibration & 13\% \\
& & ~~~\met & 9\% \\
& & ~~~Diboson modelling & 7\% \\
& & ~~~Electron calibration & 6\% \\
\hline\hline
\end{tabular}
\end{table}

\renewcommand{\arraystretch}{1.5}
\renewcommand{\arraystretch}{1.0}

The compatibility of the observed \mmc spectrum with the background-only hypothesis, for a given $A$ boson mass hypothesis, is assessed with a local $p_0$-value based on the profile-likelihood-ratio test statistic~\cite{Cowan:2010js}.
No significant deviation from the background-only hypothesis is observed.
The largest local excess of events is \SI{1.8}{$\sigma$}, observed for an $A$ boson mass of \SI{20}{\giga\electronvolt}, followed by a \SI{1.3}{$\sigma$} excess for the \SI{85}{\giga\electronvolt} mass hypothesis.

\begin{figure}[!t]
\vspace{0pt}
\centering
\includegraphics[width=0.85\textwidth]{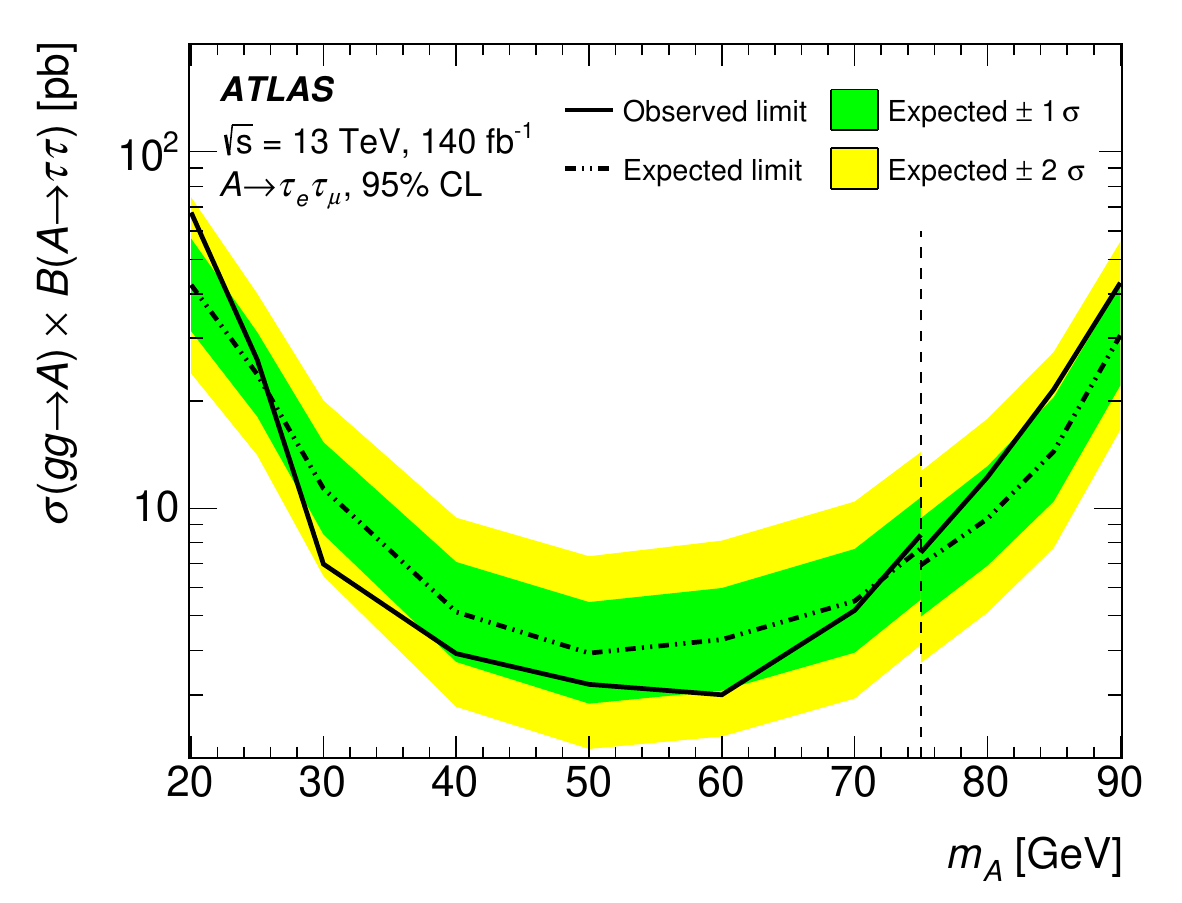}
\caption[Observed cross-section times branching ratio limit]{Expected and observed upper limits on the $gg\to A$ production cross-section times the branching ratio for $A$ boson decay into two $\tau$-leptons, for $A$ boson masses from \num{20} to \SI{90}{\giga\electronvolt}. The dash-dotted line shows the expected limit with its \SI{1}{$\sigma$} and \SI{2}{$\sigma$} uncertainty bands in green and yellow. The solid line shows the observed limit. The vertical dashed line indicates the splitting of the SR: the mass hypotheses on the left side are analysed within the low-mass SR and the ones on the right side in the high-mass SR.}
\label{fig:xsec_limit_unblinded}
\end{figure}

The observed and expected \SI{95}{\%} CL exclusion limits are calculated using a modified frequentist approach, the CL$_\text{s}$ method~\cite{Read:2002hq}, with the asymptotic approximation to the test-statistic distribution~\cite{Cowan:2010js}.
Limits are set on the signal strength $\mu$ and translated into limits on the gluon--gluon fusion production cross-section $\sigma(gg\to A)$ times the branching ratio for $A$ boson decay into two $\tau$-leptons, \bro{$A\to\tau\tau$}.
The observed and expected upper limits are shown in \cref{fig:xsec_limit_unblinded} and include all systematic uncertainties detailed in \cref{sec:systematics} except for the signal cross-section uncertainties.
The observed limits range from \SI{68}{pb} for $m_A=\SI{20}{\giga\electronvolt}$ to \SI{3.0}{pb} for $m_A=\SI{60}{\giga\electronvolt}$, increasing to \SI{43}{pb} at $m_A=\SI{90}{\giga\electronvolt}$ due to the reduced sensitivity near the $Z$-boson resonance.

\begin{figure}[!t]
\vspace{0pt}
\centering
\includegraphics[width=0.85\textwidth]{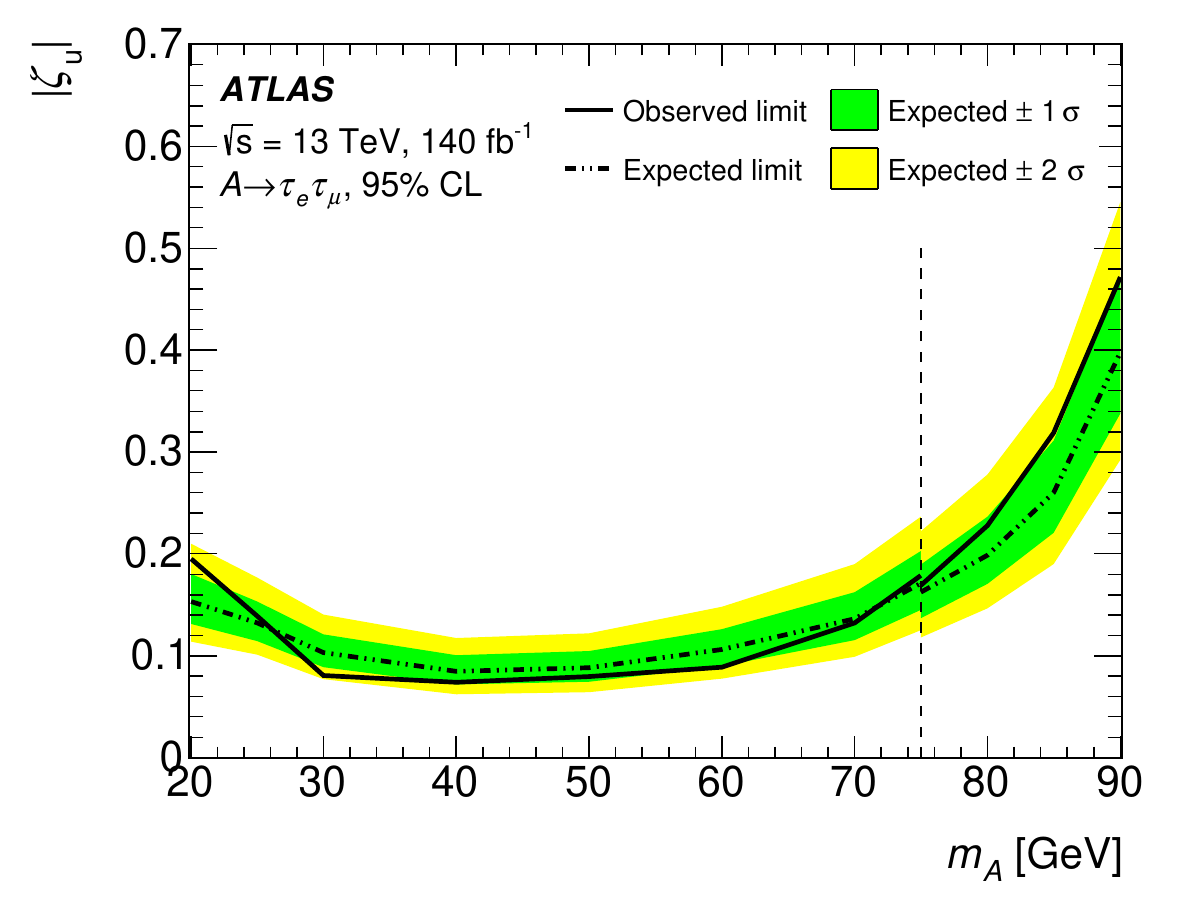}
\caption[Observed upper limit on the absolute value of the up-type quark coupling parameter $|\zeta_\mathrm{u}|$]{Expected and observed upper limits on the absolute value of the up-type quark coupling parameter \azu for masses of the $A$ boson ranging from \num{20} to \SI{90}{\giga\electronvolt}, assuming an exclusive decay of the $A$ boson into a $\tau^+\tau^-$ pair. The dash-dotted line shows the expected limit with its \SI{1}{$\sigma$} and \SI{2}{$\sigma$} uncertainty bands in green and yellow. The solid line shows the observed limit. The vertical dashed line indicates the splitting of the SR: the mass hypotheses on the left side are analysed within the low-mass SR and the ones on the right side in the high-mass SR.}
\label{fig:zeta_limit_unblinded}
\end{figure}

In addition, this analysis provides an upper limit on the absolute value of the up-type quark coupling parameter \azu within the flavour-aligned 2HDM model. For this, the uncertainties in the cross-sections are added to the fit. Only gluon--gluon fusion production of the $A$ boson via a top-quark loop is considered and an exclusive decay of the $A$ boson into a $\tau^+\tau^-$ pair with $\mathcal{B}(A\to\tau\tau)=1$ is assumed, according to the model constraints discussed in \cref{sec:introduction}. Thus, the obtained limits $\sigma_\text{fit}$ can be transformed into limits on \azu via
\begin{equation*}
|\zeta_\mathrm{u}| = \sqrt{\dfrac{\sigma_\text{fit}}{\sigma_\text{ref}}},
\end{equation*}
where $\sigma_\text{ref}$ is the cross-section calculated with \texttt{ggHiggs}~\cite{Ball_2013,Bonvini_2014,Bonvini_2016,Ahmed_2016,Bonvini_2018,Bonvini_2013,Harlander_2002,Harlander:2009mq,Harlander:2009my,anastasiou2014,Anastasiou:2016cez,Mistlberger_2018} for each of the $A$ boson mass hypotheses.
The calculated observed and expected upper limits are displayed in~\cref{fig:zeta_limit_unblinded}.
As a consequence of the higher production cross-sections for lower $A$ boson masses, the upper limits are more stringent for the lower signal-mass hypotheses around \SI{40}{\giga\electronvolt} and rise with increasing mass. The quadratic \azu dependence attenuates this effect slightly.

\FloatBarrier
\section{Conclusion}
%
%

A search for a light CP-odd Higgs boson, $A$, produced in proton--proton collisions at a centre-of-mass energy of \SI{13}{\tera\electronvolt} during \RunTwo of the LHC has been carried out, using the $\lumi$ dataset collected with the ATLAS detector.
Results are reported for the decay channels $A \to \tau^{+}\tau^{-} \to \mu^{+}\nu_{\mu} \bar{\nu}_{\tau} e^{-} \bar{\nu}_{e} \nu_{\tau}$ and $A \to \tau^{+}\tau^{-} \to e^{+}\nu_{e}\bar{\nu}_{\tau} \mu^-\bar{\nu}_{\mu}\nu_{\tau}$.
The search is performed separately in low-mass and high-mass signal regions in order to optimize the signal sensitivity.
A data-driven matrix method is applied to estimate the background of fake and non-prompt leptons.
Other background expectations are determined by MC simulations.
The \mmc variable is used as an estimator of the invariant mass of the $\tau$-lepton pair.
Its distributions in the low-mass and high-mass signal regions are utilised in the statistical interpretation of the data.
No significant excess of events above the SM background prediction is observed.
The \SI{95}{\%} CL upper limits placed on the gluon--gluon fusion production cross-section for a CP-odd Higgs boson decaying into the $\tau^+\tau^-$ final state vary between \SI{3.0}{pb} and \SI{68}{pb} in the $A$ boson mass range between \SI{20}{\giga\electronvolt} and \SI{90}{\giga\electronvolt}. The mass range below \SI{60}{\giga\electronvolt} had not been probed before.
An interpretation of the results in the context of the flavour-aligned two-Higgs-doublet model is also given.
Upper limits between 0.074 and 0.47 at \SI{95}{\%} CL are set on the absolute value of the up-type quark coupling parameter \azuu, improving on previous limits on this parameter~\cite{PhysRevD.98.035001}.


%
\section*{Acknowledgements}
%

%
%

%
%

We thank CERN for the very successful operation of the LHC and its injectors, as well as the support staff at
CERN and at our institutions worldwide without whom ATLAS could not be operated efficiently.

The crucial computing support from all WLCG partners is acknowledged gratefully, in particular from CERN, the ATLAS Tier-1 facilities at TRIUMF/SFU (Canada), NDGF (Denmark, Norway, Sweden), CC-IN2P3 (France), KIT/GridKA (Germany), INFN-CNAF (Italy), NL-T1 (Netherlands), PIC (Spain), RAL (UK) and BNL (USA), the Tier-2 facilities worldwide and large non-WLCG resource providers. Major contributors of computing resources are listed in Ref.~\cite{ATL-SOFT-PUB-2023-001}.

We gratefully acknowledge the support of ANPCyT, Argentina; YerPhI, Armenia; ARC, Australia; BMWFW and FWF, Austria; ANAS, Azerbaijan; CNPq and FAPESP, Brazil; NSERC, NRC and CFI, Canada; CERN; ANID, Chile; CAS, MOST and NSFC, China; Minciencias, Colombia; MEYS CR, Czech Republic; DNRF and DNSRC, Denmark; IN2P3-CNRS and CEA-DRF/IRFU, France; SRNSFG, Georgia; BMBF, HGF and MPG, Germany; GSRI, Greece; RGC and Hong Kong SAR, China; ISF and Benoziyo Center, Israel; INFN, Italy; MEXT and JSPS, Japan; CNRST, Morocco; NWO, Netherlands; RCN, Norway; MNiSW, Poland; FCT, Portugal; MNE/IFA, Romania; MSTDI, Serbia; MSSR, Slovakia; ARIS and MVZI, Slovenia; DSI/NRF, South Africa; MICIU/AEI, Spain; SRC and Wallenberg Foundation, Sweden; SERI, SNSF and Cantons of Bern and Geneva, Switzerland; NSTC, Taipei; TENMAK, T\"urkiye; STFC/UKRI, United Kingdom; DOE and NSF, United States of America.

Individual groups and members have received support from BCKDF, CANARIE, CRC and DRAC, Canada; CERN-CZ, FORTE and PRIMUS, Czech Republic; COST, ERC, ERDF, Horizon 2020, ICSC-NextGenerationEU and Marie Sk{\l}odowska-Curie Actions, European Union; Investissements d'Avenir Labex, Investissements d'Avenir Idex and ANR, France; DFG and AvH Foundation, Germany; Herakleitos, Thales and Aristeia programmes co-financed by EU-ESF and the Greek NSRF, Greece; BSF-NSF and MINERVA, Israel; NCN and NAWA, Poland; La Caixa Banking Foundation, CERCA Programme Generalitat de Catalunya and PROMETEO and GenT Programmes Generalitat Valenciana, Spain; G\"{o}ran Gustafssons Stiftelse, Sweden; The Royal Society and Leverhulme Trust, United Kingdom.

In addition, individual members wish to acknowledge support from Armenia: Yerevan Physics Institute (FAPERJ); CERN: European Organization for Nuclear Research (CERN PJAS); Chile: Agencia Nacional de Investigaci\'on y Desarrollo (FONDECYT 1230812, FONDECYT 1230987, FONDECYT 1240864); China: Chinese Ministry of Science and Technology (MOST-2023YFA1605700), National Natural Science Foundation of China (NSFC - 12175119, NSFC 12275265, NSFC-12075060); Czech Republic: Czech Science Foundation (GACR - 24-11373S), Ministry of Education Youth and Sports (FORTE CZ.02.01.01/00/22\_008/0004632), PRIMUS Research Programme (PRIMUS/21/SCI/017); EU: H2020 European Research Council (ERC - 101002463); European Union: European Research Council (ERC - 948254, ERC 101089007), European Union, Future Artificial Intelligence Research (FAIR-NextGenerationEU PE00000013), Italian Center for High Performance Computing, Big Data and Quantum Computing (ICSC, NextGenerationEU); France: Agence Nationale de la Recherche (ANR-20-CE31-0013, ANR-21-CE31-0013, ANR-21-CE31-0022, ANR-22-EDIR-0002); Germany: Baden-Württemberg Stiftung (BW Stiftung-Postdoc Eliteprogramme), Deutsche Forschungsgemeinschaft (DFG - 469666862, DFG - CR 312/5-2); Italy: Istituto Nazionale di Fisica Nucleare (ICSC, NextGenerationEU), Ministero dell'Università e della Ricerca (PRIN - 20223N7F8K - PNRR M4.C2.1.1); Japan: Japan Society for the Promotion of Science (JSPS KAKENHI JP22H01227, JSPS KAKENHI JP22H04944, JSPS KAKENHI JP22KK0227, JSPS KAKENHI JP23KK0245); Netherlands: Netherlands Organisation for Scientific Research (NWO Veni 2020 - VI.Veni.202.179); Norway: Research Council of Norway (RCN-314472); Poland: Ministry of Science and Higher Education (IDUB AGH, POB8, D4 no 9722), Polish National Agency for Academic Exchange (PPN/PPO/2020/1/00002/U/00001), Polish National Science Centre (NCN 2021/42/E/ST2/00350, NCN OPUS 2023/51/B/ST2/02507, NCN OPUS nr 2022/47/B/ST2/03059, NCN UMO-2019/34/E/ST2/00393, NCN \& H2020 MSCA 945339, UMO-2020/37/B/ST2/01043, UMO-2021/40/C/ST2/00187, UMO-2022/47/O/ST2/00148, UMO-2023/49/B/ST2/04085, UMO-2023/51/B/ST2/00920); Slovenia: Slovenian Research Agency (ARIS grant J1-3010); Spain: Generalitat Valenciana (Artemisa, FEDER, IDIFEDER/2018/048), Ministry of Science and Innovation (MCIN \& NextGenEU PCI2022-135018-2, MICIN \& FEDER PID2021-125273NB, RYC2019-028510-I, RYC2020-030254-I, RYC2021-031273-I, RYC2022-038164-I); Sweden: Carl Trygger Foundation (Carl Trygger Foundation CTS 22:2312), Swedish Research Council (Swedish Research Council 2023-04654, VR 2018-00482, VR 2022-03845, VR 2022-04683, VR 2023-03403, VR grant 2021-03651), Knut and Alice Wallenberg Foundation (KAW 2018.0458, KAW 2019.0447, KAW 2022.0358); Switzerland: Swiss National Science Foundation (SNSF - PCEFP2\_194658); United Kingdom: Leverhulme Trust (Leverhulme Trust RPG-2020-004), Royal Society (NIF-R1-231091); United States of America: U.S. Department of Energy (ECA DE-AC02-76SF00515), Neubauer Family Foundation.

%
%


%
%
%
%
%
%
%

%
\clearpage
\appendix
\printbibliography
\clearpage
 
\begin{flushleft}
\hypersetup{urlcolor=black}
{\Large The ATLAS Collaboration}

\bigskip

\AtlasOrcid[0000-0002-6665-4934]{G.~Aad}$^\textrm{\scriptsize 105}$,
\AtlasOrcid[0000-0001-7616-1554]{E.~Aakvaag}$^\textrm{\scriptsize 17}$,
\AtlasOrcid[0000-0002-5888-2734]{B.~Abbott}$^\textrm{\scriptsize 124}$,
\AtlasOrcid[0000-0002-0287-5869]{S.~Abdelhameed}$^\textrm{\scriptsize 120a}$,
\AtlasOrcid[0000-0002-1002-1652]{K.~Abeling}$^\textrm{\scriptsize 57}$,
\AtlasOrcid[0000-0001-5763-2760]{N.J.~Abicht}$^\textrm{\scriptsize 51}$,
\AtlasOrcid[0000-0002-8496-9294]{S.H.~Abidi}$^\textrm{\scriptsize 30}$,
\AtlasOrcid[0009-0003-6578-220X]{M.~Aboelela}$^\textrm{\scriptsize 46}$,
\AtlasOrcid[0000-0002-9987-2292]{A.~Aboulhorma}$^\textrm{\scriptsize 36e}$,
\AtlasOrcid[0000-0001-5329-6640]{H.~Abramowicz}$^\textrm{\scriptsize 156}$
% ...
\bigskip
\\

% ...
$^{17}$Department for Physics and Technology, University of Bergen, Bergen; Norway.\\
% ...
$^{30}$Physics Department, Brookhaven National Laboratory, Upton NY; United States of America.\\
% ...
$^{36}$$^{(a)}$Facult\'e des Sciences Ain Chock, Universit\'e Hassan II de Casablanca;$^{(b)}$Facult\'{e} des Sciences, Universit\'{e} Ibn-Tofail, K\'{e}nitra;$^{(c)}$Facult\'e des Sciences Semlalia, Universit\'e Cadi Ayyad, LPHEA-Marrakech;$^{(d)}$LPMR, Facult\'e des Sciences, Universit\'e Mohamed Premier, Oujda;$^{(e)}$Facult\'e des sciences, Universit\'e Mohammed V, Rabat;$^{(f)}$Institute of Applied Physics, Mohammed VI Polytechnic University, Ben Guerir; Morocco.\\
% ...
$^{46}$Physics Department, Southern Methodist University, Dallas TX; United States of America.\\
% ...
$^{51}$Fakult\"{a}t Physik , Technische Universit{\"a}t Dortmund, Dortmund; Germany.\\
% ...
$^{57}$II. Physikalisches Institut, Georg-August-Universit\"{a}t G\"ottingen, G\"ottingen; Germany.\\
% ...
$^{105}$CPPM, Aix-Marseille Universit\'e, CNRS/IN2P3, Marseille; France.\\
% ...
$^{120}$$^{(a)}$New York University Abu Dhabi, Abu Dhabi;$^{(b)}$United Arab Emirates University, Al Ain; United Arab Emirates.\\
% ...
$^{124}$Homer L. Dodge Department of Physics and Astronomy, University of Oklahoma, Norman OK; United States of America.\\
% ...
$^{156}$Raymond and Beverly Sackler School of Physics and Astronomy, Tel Aviv University, Tel Aviv; Israel.\\
% ...
$^{a}$ Also Affiliated with an institute covered by a cooperation agreement with CERN.\\
$^{b}$ Also at An-Najah National University, Nablus; Palestine.\\
$^{c}$ Also at Borough of Manhattan Community College, City University of New York, New York NY; United States of America.\\
$^{d}$ Also at Center for High Energy Physics, Peking University; China.\\
$^{e}$ Also at Center for Interdisciplinary Research and Innovation (CIRI-AUTH), Thessaloniki; Greece.\\
$^{f}$ Also at CERN, Geneva; Switzerland.\\
% ...

\end{flushleft}


%

%
%

%

%
%
%
%

\end{document}